\documentclass[10 pt,final,journal,letterpaper,oneside,twocolumn]{IEEEtran}
\usepackage{multicol}   
\usepackage{multirow}
\usepackage[pdftex]{graphicx}
\usepackage[ruled]{algorithm2e}
\usepackage{algorithmic}
\usepackage[small]{caption}
\usepackage{float}
\usepackage{amssymb,amsmath}
\usepackage{graphicx}
\usepackage{subfigure}
\usepackage{xspace}
\usepackage{amsthm}
\usepackage{fancyhdr}

\usepackage{caption,setspace}
\usepackage{amssymb,amsmath}
\usepackage{bm}
\usepackage{textcomp} 
\usepackage{epstopdf}
\usepackage{bbding}
\usepackage{cite} 
\usepackage{color}
\usepackage{pifont}
\usepackage{lipsum}
\usepackage{amstext}

\usepackage{cuted}

\graphicspath{{fig/}}

\newcommand{\g} {\mathbf{g}}

\newcommand\scalemath[2]{\scalebox{#1}{\mbox{\ensuremath{\displaystyle #2}}}}

\begin{document}

\title{\huge Energy-Efficient Beamforming and Resource Optimization for AmBSC-Assisted Cooperative NOMA IoT Networks}
\author{Muhammad Asif, Asim Ihsan, Wali Ullah Khan,  \IEEEmembership{Member, IEEE}, Ali Ranjha, Shengli Zhang, \IEEEmembership{Senior Member, IEEE}, and Sissi Xiaoxiao Wu, \IEEEmembership{Member, IEEE}, \thanks{Muhammad Asif, Shengli Zhang, and Sissi Xiaoxiao Wu are with the Guangdong Key Laboratory of Intelligent Information Processing, College of Electronics and Information Engineering, Shenzhen University, Shenzhen, Guangdong 518060, China. (emails: masif@szu.edu.cn, zsl@szu.edu.cn, xxwu.eesissi@gmail.com).

Asim Ihsan is with the School of Computer Science and Electronic Engineering, Bangor University, Bangor LL57 1UT, U.K. (e-mail:
a.ihsan@bangor.ac.uk)
		
Wali Ullah Khan is with the Interdisciplinary Centre for Security, Reliability and Trust (SnT), University of Luxembourg, 1855 Luxembourg City, Luxembourg (Emails: waliullah.khan@uni.lu, waliullahkhan30@gmail.com).

Ali Ranjha is with the Department of Electrical Engineering, École de Technologie Supérieure, Montréal, Quebec, Canada, (email: ali- nawaz.ranjha.1@ens.estmtl.ca).

(Corresponding author: Sissi Xiaoxiao Wu.)
}

\vspace{-0.6cm}}%

\markboth{IEEE Internet of Things Journal}%
{Shell \MakeLowercase{\textit{et al.}}: Bare Demo of IEEEtran.cls for IEEE Journals} 

\maketitle

\begin{abstract}
In this manuscript, we present an energy-efficient alternating optimization framework based on the multi-antenna ambient backscatter communication (AmBSC) assisted cooperative non-orthogonal multiple access (NOMA) for next-generation (NG) internet-of-things (IoT) enabled communication networks. Specifically, the energy-efficiency maximization is achieved for the considered AmBSC-enabled multi-cluster cooperative IoT NOMA system by optimizing the active-beamforming vector and power-allocation coefficients (PAC) of IoT NOMA users at the transmitter, as well as passive-beamforming vector at the multi-antenna assisted backscatter node. Usually, increasing the number of IoT NOMA users in each cluster results in inter-cluster interference (ICI) (among different clusters) and intra-cluster interference (among IoT NOMA users). To combat the impact of ICI, we exploit a zero-forcing (ZF) based active-beamforming, as well as an efficient clustering technique at the source node. Further, the effect of intra-cluster interference is mitigated by exploiting an efficient power-allocation policy that determines the PAC of IoT NOMA users under the quality-of-service (QoS), cooperation, SIC decoding, and power-budget constraints. Moreover, the considered non-convex passive-beamforming problem is transformed into a standard semi-definite programming (SDP) problem by exploiting the successive-convex approximation (SCA) approximation, as well as the difference of convex (DC) programming, where Rank-1 solution of passive-beamforming is obtained based on the penalty-based method. Furthermore, the numerical analysis of simulation results demonstrates that the proposed energy-efficiency maximization algorithm exhibits an efficient performance by achieving convergence within only a few iterations.      
   
\end{abstract}


\IEEEpeerreviewmaketitle

\section{Introduction}
\IEEEPARstart{F} {uture} next-generation (NG) internet-of-things (IoT) networks are expected to face different challenges due to limited energy and spectrum resources to support the connectivity of billions of IoT devices. Therein, the maintenance of these IoT nodes could be very challenging, time-consuming, and expensive. Hence, it is highly desired to realize the future communication networks by deploying energy-efficient IoT nodes, because, it would be uneconomical to replace their batteries over a regular duration of time \cite{giordani2020toward}. Specifically, to fulfill the spectrum and energy requirements, some spectrum and energy-efficient technologies could be coupled with next-generation (NG) IoT networks to support the connectivity of the billions of IoT devices. In this regard, non-orthogonal multiple access (NOMA) and backscatter communication (BSC) seems to be a promising solution to realize the sustainable future wireless networks by deploying a large number of low-cost energy-efficient IoT devices \cite{jameel2020noma,liu2019next}. 

The backscatter communication has been included in the list of potential competitors for future energy-efficient communication networks by providing a flexible scale of attributes including low-cost, ultra-low power consumption (less than $1$ $\mu W$), and almost zero maintenance \cite{liu2022covert}. Therefore, backscatter communication has grabbed a tremendous attention of the research community, working for beyond fifth-generation (5G) communication protocols, across the academic and industrial zones. Typically, in conventional backscatter communication, such as radio-frequency identification (RFID), the RFID transceiver provides a continuous-wave carrier signal to power up the backscatter tag which reflects the modulated signal towards RFID receiver \cite{yang2020exploiting}. Recently, a new protocol in the domain of backscatter communication, known as ambient backscatter communication (AmBSC), has been considered as a new technological revolution for beyond 5G communication systems \cite{wang2021coherent,jameel2020low}. Unlike conventional backscatter communication, the ambient backscatter tag harvests the energy from radio signals available in the surrounding environment to fulfill its energy requirements. Then, the ambient backscatter tag uses the harvested power to reflect the modulated signal towards the desired nodes in the network. Moreover, the ambient backscattering could be an anomalous solution to extend the coverage of the network when the direct link between access point (AP) and IoT receiver is blocked by the obstacles. Base on the aforementioned merits of backscatter communication, several research efforts have been devoted to investigate the different aspects of the backscatter-enabled wireless communication networks \cite{luan2021better, guo2018exploiting, huang2020freescatter,li2020hybrid,he2020monostatic, he2020simple,wu2021beamforming,ma2020joint, devineni2018ambient,liu2020deep,tao2020optimal,hu2021performance,han2021uav}. 

Similarly, non-orthogonal multiple access (NOMA) has been recognized as a potential competitor among different multiple access techniques in enabling sustainable spectrum-efficient beyond 5G IoT-enabled communication networks \cite{khan2019joint,liu2022evolution,mu2021intelligent}. Unlike conventional orthogonal multiple access (OMA) schemes where the orthogonality is achieved in time/frequency domain, NOMA accommodates the multiple IoT users over the same frequency/code resource. Furthermore, NOMA outperforms its OMA counterparts in terms of spectral efficiency, energy efficiency, and massive connectivity, where multiplexing is achieved by assigning different power levels to IoT users. Different from OMA, the NOMA protocol provides fairness among different IoT users where more power is allocated to the distant IoT user (far user) with bad channel conditions, and less power is assigned to the IoT user close to the AP (near user) with good channel condition. The practical realization of NOMA is achieved by two prominent technologies, known as superposition coding (SC) and successive interference cancellation (SIC), at the transmitter and receiver, respectively \cite{makki2020survey,ding2017application}.

\subsection{Technical Literature Review}
The efficient utilization of available resources, in terms of spectrum and energy, has become the primary interest of research for the NG-IoT networks to support a huge number of network IoT devices. Based on the aforementioned discussion, the integration of NOMA and AmBSC could play a crucial role in the development of future IoT networks to fulfill the requirements in terms of capacity, spectral efficiency, energy-efficiency, coverage, and massive connectivity \cite{khan2021noma,tanveer2021enhanced,jameel2019applications}.

1) {\it Backscatter Communication for Conventional NOMA Networks:} Authors in \cite{zhang2019backscatter}, have computed the closed-form expressions for ergodic channel capacity and outage-probability (OP) for ambient backscatter-enabled cellular networks. Nazar {\em et al.} \cite{nazar2021ber} have investigated the bit-error-rate (BER) performance of the backscatter-enabled NOMA system by calculating the closed-form expressions. Besides, in \cite{li2020secrecy}, the analytical expressions for OP intercept-probability (IP) have been investigated along with the security problems for ambient backscatter-enabled NOMA communication system. Also, In \cite{li2021hardware}, physical layer security of backscatter-enabled NOMA in terms of reliability and security is investigated. Further, Khan {\em et al.} \cite{khan2021backscatter123} proposed an optimization framework by exploiting a backscatter-enabled NOMA system for vehicle-to-everything (V2X) communication protocol for optimizing the power allocation of the transmitter. An efficient optimization framework to optimize the power allocation at the transmitter node and reflection-coefficient at the backscatter tag has been presented in \cite{khan2021backscatter}. Authors in \cite{xu2020energy} have proposed an optimization framework to maximize the energy efficiency (EE) of the backscatter-enabled NOMA protocol. Furthermore, an optimization framework for optimizing the power budget, power allocation coefficients, and reflection coefficient of ambient backscatter tag has been investigated for backscatter NOMA-enabled internet-of-vehicle (IoV) network\cite{khan2021energy}.  Authors of \cite{chen2021backscatter}, an optimization framework is investigated that simultaneously optimizes the power allocation to maximize the ergodic sum capacity for a NOMA-enabled backscatter communication network. Further, in \cite{khan25}, an optimization framework that optimizes the power allocation at the source and reflection-coefficient at the backscatter tag is proposed to minimize the total transmit power for backscatter-enabled V2X communication under NOMA protocol. 

2) {\it Backscatter Communication for IoT-enabled NOMA Networks}: Of late, several research studies have been conducted to exploit the backscatter-enabled NOMA for NG-IoT networks. For example, in \cite{yang2019resource}, Yang {\em et al.} proposed an optimization framework by optimizing the time and reflection coefficient of backscatter for a backscatter NOMA-enabled  IoT network. The secrecy rate of the backscatter-NOMA IoT system has been maximized by optimizing the reflection coefficient of the backscatter tag under the multi-cell scenario in \cite{khan2020secure}. Authors in \cite{li2021physical}, investigated the security and reliability, in terms of OP and IP, of AmBSC-enabled green IoT network. In \cite {le2021joint}, the closed-form expressions, in terms of OP and ergodic sum rate, have been computed for the AmBSC-enabled IoT NOMA system. Besides, in \cite{khan2021joint}, the ergodic sum capacity of a backscatter-enabled IoT NOMA system has been maximized by jointly optimizing the transmit power at the AP node and reflection-coefficient at the backscatter tag. Moreover,  Sacarelo {\em et al.}\cite{sacarelo2021beamforming} proposed an optimization framework to achieve maximum fairness by optimizing the transmit beamforming, received beamforming, and reflection-coefficients of backscatter nodes for NOMA-enabled IoT network. The physical layer security of an AmBSC-enabled IoT NOMA network has been improved by maximizing the secrecy rate of the system \cite{khan2022joint}. Finally, Ahmed {\em et al.}\cite{ahmed2021backscatter} an alternating optimization framework has been proposed to maximize the EE of backscatter-assisted multi-cell IoT network by optimizing the transmit power of AP, PAC, and reflection-coefficients of backscatter nodes.

\subsection{Motivation and Contributions}
In the aforementioned literature \cite{zhang2019backscatter} -- \cite{khan25}, the performance of the system has been demonstrated by integrating NOMA with the backscatter communication. However, cooperation among the NOMA users is not considered. Besides, all of these works have considered only a single cluster which is not practical for the actual realization of BSC-enabled NOMA communication. Moreover, only a single antenna is considered at the backscatter tag, where the performance of the system is improved by optimizing the single-antenna reflection coefficient at the backscatter tag (BST). Further, the transmit-beamforming has not been considered in the existing studies on backscatter-assisted NOMA networks. Accordingly, in literature \cite{yang2019resource} -- \cite{ahmed2021backscatter}, the backscatter communication is integrated with NOMA protocol to realize the different IoT networks. However, the optimization of active passive beamforming vectors has not been considered at the source and backscatter nodes, respectively. Moreover, please note that these works also do not consider the cooperation among IoT NOMA users to improve the performance of the system.         

The proliferation of a large number of IoT devices for NG-IoT networks would require an efficient utilization of available resources to support the connectivity of a huge number of network IoT devices. Based on the above discussion and technical literature review, it would be very interesting to investigate the AmBSC-assisted cooperative IoT NOMA system under a multi-cluster scenario, where the transmitter and backscatter nodes are equipped with multiple antennas. Specifically, the active and passive NOMA-beamforming can be achieved at the AP and backscatter tag, equipped with multiple antennas, respectively. Also, NOMA-beamforming design could be very helpful to enhance the performance of multi-cluster backscatter-enabled cooperative NOMA IoT system, where the IoT NOMA users in each cluster can be served by a common beamformer through NOMA protocol. To the best of our knowledge, an optimization framework that simultaneously optimizes the power-allocation coefficients (PAC) and zero-forcing (ZF) based transmit-beamforming vector at the source node, as well as the passive-beamforming vector at the multi-antenna assisted backscatter tag for a backscatter-assisted multi-cluster cooperative NOMA-enabled IoT network has not been investigated yet. The major contributions of this work are summarized as follows:

\begin{enumerate} 
\item An energy-efficient AmBSC-enabled cooperative-NOMA system has been considered to realize a multi-cluster IoT network, where both transmitter and backscatter nodes are equipped with multiple antennas. The performance of the considered IoT system has been improved, in terms of energy-efficiency, by active and passive beam-forming vectors at AP and backscatter node, respectively. Moreover, cooperation among IoT NOMA users has been exploited to improve the fairness and performance of the far IoT NOMA user.  

\item Further, we propose an energy-efficient optimization framework that maximizes the energy-efficiency for the considered backscatter-enabled cooperative IoT NOMA system by simultaneously optimizing the power-allocation coefficient (PAC) and active-beamforming vector at the AP, as well as passive-beamforming vector at the multi-antenna backscatter tag under the quality-of-service (QoS), SIC decoding, cooperation, reflection-coefficients, and power budget constraints.   

\item The proposed alternating optimization algorithm has been decoupled into two stages. In the first stage, ZF-based active-beamforming and PAC are computed. Therein, ZF-based active-beamforming along with efficient clustering has been exploited to mitigate the effect of inter-cluster interference (ICI). Subsequently, the intra-cluster interference has been mitigated by efficiently computing the power-allocation coefficients for IoT NOMA users in each cluster based on the successive-convex approximation (SCA), Dinkelbach method, and dual theory followed by the sub-gradient method. In the second stage, the considered non-convex passive-beamforming problem is transformed into a standard semi-definite programming (SDP) problem by exploiting the SCA approximation, as well as the difference of convex (DC) programming, where Rank-1 solution of passive-beamforming has been computed based on the penalty-based method.

\item Finally, the numerical simulation results demonstrate that the proposed alternating optimization framework provides an efficient performance, in terms of energy-efficiency, to realize the considered AmBSC-assisted cooperative-NOMA IoT network. Moreover, the proposed energy-efficient optimization algorithm exhibits low complexity and convergence within a few iterations.  
\end{enumerate}

The remainder of this work is arranged as follows: In Section II, the system model for considered BST-enabled cooperative NOMA for multi-clustered IoT system along with the problem formulation for the energy-efficiency maximization is provided. The optimal solution to considered energy-efficiency maximization problem is provided in Section III. Further, in Section IV, the demonstration of numerical simulation results is provided. Finally, Section V provides the conclusion of the proposed work.
\begin{figure*}[!t]
\centering
\includegraphics [width=0.70\textwidth]{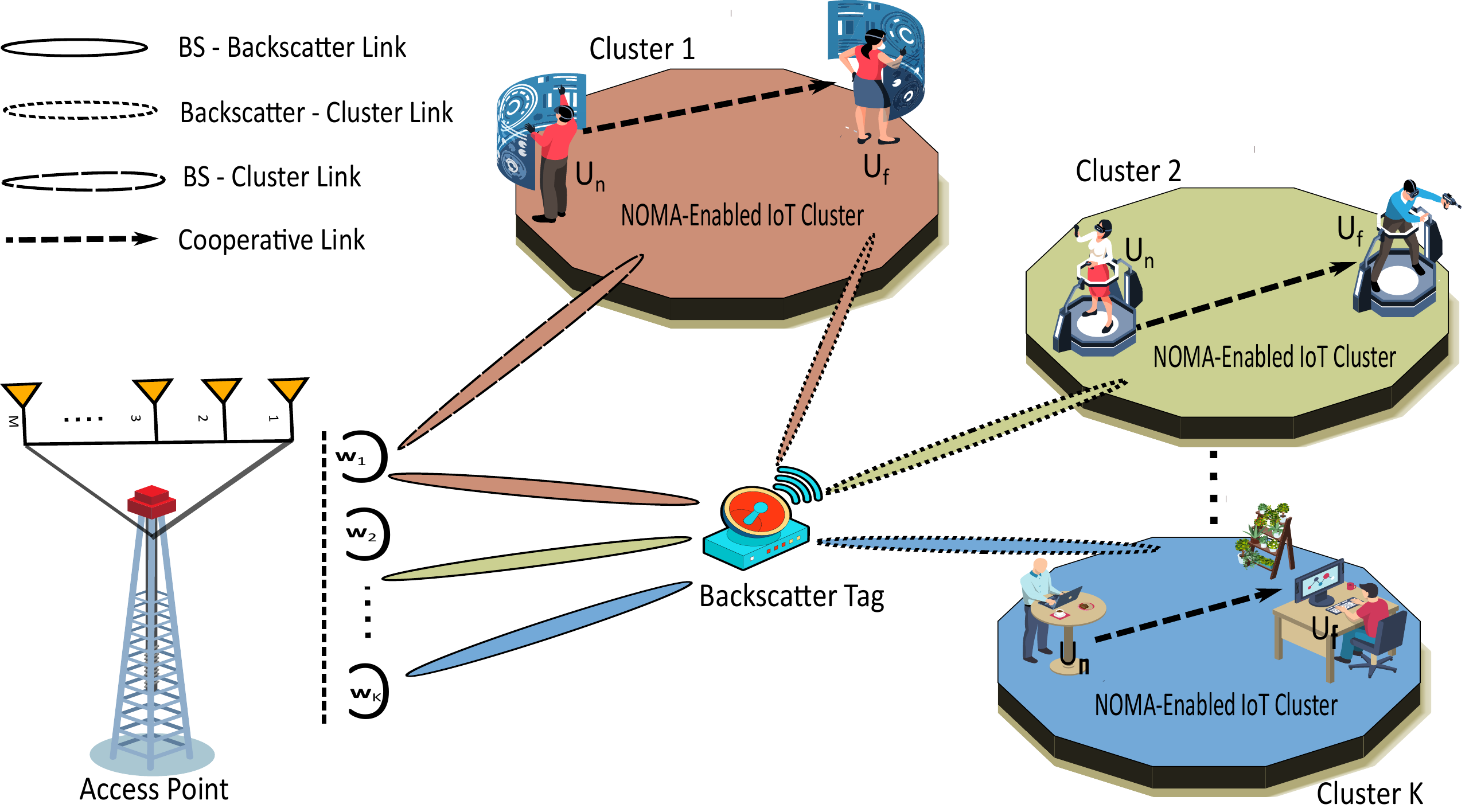}
\caption{System Model for the Considered AmBSC-Assisted multi-cluster cooperative IoT NOMA Network}
\label{blocky}
\end{figure*}
\section{Proposed System Model and Problem Formulation}
In this work, a multi-antenna AmBSC-assisted cooperative IoT NOMA-beamforming system has been depicted in Fig. \ref{blocky}, where an access point (AP) is equipped with $M$ antennas having uniform linear array (ULA) response, and an ambient backscatter tag realized by $N$ antennas with ULA response. Hence, the backscatter node assists to improve the performance of the NOMA-enabled single-antenna IoT users in $K$ clusters. For the convenience of analysis, it is assumed that each cluster consists of two IoT NOMA users. According to the NOMA principle, the IoT user close to the AP, in the $k^{th}$ cluster, with good channel condition is labeled as near IoT user $U^k_{n}$, whereas, the IoT user with bad channel condition, known as far IoT user, is denoted as $U^k_{f}$. Further, it is assumed that the perfect channel-state information (CSI) is known at the AP node. Hence, AP can transmit with $K$ beamforming vectors to accommodate $K$ clusters in a considered downlink backscatter-enabled cooperative NOMA IoT system, where the IoT users in each cluster are served by a common beamformer based on the NOMA protocol.

It is important to note that more IoT users can be served by the proposed NOMA-beamforming system, but, it results in intra-cluster interference among IoT NOMA users along with the ICI between different clusters. Therefore, to enhance the performance of the IoT NOMA users in each cluster, efficient interference management must be incorporated to mitigate the effect of intra-cluster interference as well as ICI. In this scenario, ZF-based beamforming, as well as an efficient clustering technique based on the maximum correlation and channel gain difference among the IoT NOMA users has been exploited to mitigate the effect of ICI. Subsequently, an efficient power-allocation policy has been proposed, to mitigate the effect of intra-cluster interference, which computes the PAC of IoT NOMA users in each cluster of the considered system. Moreover, a more practical geometric channel model has been adopted to demonstrate the performance of the considered system. Therein, for a direct link between AP and IoT NOMA users, the Rayleigh-fading transmission is considered because the line-of-sight (LOS) path could be blocked by the obstacles. However, for the indirect link between AP and IoT NOMA users through backscatter tag, Rician-fading transmission has been assumed, where the line-of-sight components depend upon the uniform linear array (ULA) response at the AP and multi-antenna backscatter tag while the non-line-of-sight (NLOS) components are modeled as complex Gaussian random variables having zero mean and unit variance.

\subsection{Direct Transmission Phase}
In the first time frame for considered AmBSC-assisted multi-cluster cooperative IoT NOMA system, the signals received at $U^k_{n}$ and $U^k_{f}$ in the $k^{th}$ cluster are given as follows    
      
\begin{align}
	y^{k}_{n,1}&=({\mathbf {f}^k_{n}}^H+{\mathbf {h}^k_{n}}^H\mathbf{G}\mathbf{H}^k_n)\mathbf{w}_k s_k m_k\nonumber\\
	&+({\mathbf {f}^k_{n}}^H+{\mathbf {h}^k_{n}}^H\mathbf{G}\mathbf{H}^k_n) \sum\limits_{{\substack{l=1 \\ l\neq k}}}^K\mathbf{w}_l s_l m_l+n^k_{n,1}, \label{1}
\end{align}
and
\begin{align}
	y^{k}_{f,1}=&({\mathbf {f}^k_{f}}^H+{\mathbf {h}^k_{f}}^H\mathbf{G}\mathbf{H}^k_f)\mathbf{w}_k s_k m_k\nonumber\\
	&+({\mathbf {f}^k_{f}}^H+{\mathbf {h}^k_{f}}^H\mathbf{G}\mathbf{H}^k_f)\sum\limits_{{\substack{l=1 \\ l\neq k}}}^K\mathbf{w}_l s_l m_l+n^k_{f,1}, \label{2}
\end{align}
where $\mathbf{w}_k$ denotes the NOMA beamforming vector serving the IoT users in $k^{th}$ cluster. Also, $s_k =  \sqrt{P_k\alpha_{k,n}}x_{k,n}+ \sqrt{P_k\alpha_{k,f}}x_{k,f}$ represents the superimposed signal for NOMA receivers in the $k^{th}$ cluster, where $x_{k,n}$ and $x_{k,f}$ are the information signals intended for $U^k_{n}$ and $U^k_{f}$ with $\mathbb{E}[|x_{k,n}|^2]=1$, $\mathbb{E}[|x_{k,f}|^2]=1$, respectively, and $P_k$ represents the the transmit power assigned to $k^{th}$ cluster. Further, $\alpha_{k,n}$ and $\alpha_{k,f}$ are the power-allocation coefficients for $U^k_{n}$ and $U^k_{f}$, respectively. In addition, $m_k$ is the signal added by the backscatter tag with  $\mathbb{E}[|m_{k}|^2]=1$. Let, $\mathbf{G} =diag(g_1,g_2,\ldots,g_n)$ denotes reflection coefficient matrix (passive-beamforming matrix) of multi-antenna backscatter tag, where $g_n$ represents the reflection coefficient at the $n^{th}$ antenna of BST, whereas, $n_{n,k}$ and $n_{f,k}$ denote the additive-white Gaussian noise (AWGN) at $U^k_n$ and $U^k_f$, respectively, with zero-mean and variance $\sigma^{2}$. 

The channel gains for direct the communication from AP to $U^k_{n}$ and $U^k_{f}$, denoted as $\mathbf {f}^k_{n}$ and $\mathbf {f}^k_{f}$, are independent and identically distributed (i.i.d) complex Gaussian random variables with zero-mean and unit-variance, whereas $\mathbf {f}^k_{n}$ and $\mathbf {f}^k_{f}$ are modeled as Rayleigh-fading. Moreover, the channel gains from AP to BST for $U^k_{n}$ and $U^k_{f}$ in $k^{th}$ cluster, denoted as $\mathbf{H}^k_n$ and $\mathbf{H}^k_f$, are given as follows 
 
\begin{align}
	&\mathbf{H}^k_n=\sqrt{\eta_{0}\Big(\frac{d_{s,b}}{d_0}\Big)^{-\mu_{s,b}}}\overline{\mathbf{H}}^k_n, \label{3}
\end{align}
and

\begin{align}
	&\mathbf{H}^k_f=\sqrt{\eta_{0}\Big(\frac{d_{s,b}}{d_0}\Big)^{-\mu_{s,b}}}\overline{\mathbf{H}}^k_f, \label{4}
\end{align}
where $\eta_{0}$ represents the path loss at the reference distance $d_{0}=1$ (m). $d_{s,b}$ and $\mu_{s,b}$ denote the distance and path-loss exponent from AP to BST, respectively. Further, $\overline{\mathbf{H}}^k_n$ and $\overline{\mathbf{H}}^k_f$ are assumed under the Rician fading given as follows 

 \begin{align}
 	&	\overline{\mathbf{H}}^k_n = \sqrt{\frac{\beta_{1}}{1+\beta_{1}}}{{\mathbf{H}}^k_n}^{(LoS)}+\sqrt{\frac{1}{1+\beta_{1}}}{{\mathbf{H}}^k_n}^{(NLoS)}, \label{5}
 \end{align}
 and
 
  \begin{align}
 	&	\overline{\mathbf{H}}^k_f = \sqrt{\frac{\beta_{2}}{1+\beta_{2}}}{{\mathbf{H}}^k_f}^{(LoS)}+\sqrt{\frac{1}{1+\beta_{2}}}{{\mathbf{H}}^k_f}^{(NLoS)}, \label{6}
 \end{align}    
where ${{\mathbf{H}}^k_n}^{(LoS)} \in \mathbb{C}^{N\times M}$ and ${{\mathbf{H}}^k_f}^{(LoS)} \in \mathbb{C}^{N\times M}$ denote the LOS components from AP to BST, whereas, $\beta_{1}$ and $\beta_{2}$ are the Rician factors. ${{\mathbf{H}}^k_n}^{(NLoS)} \in \mathbb{C}^{N\times M}$, ${{\mathbf{H}}^k_f}^{(NLoS)} \in \mathbb{C}^{N\times M}$ represent the NLOS components which are i.i.d complex Gaussian random variables with mean and variance values equal to zero and one, respectively. 

\begin{figure*}
	\begin{align}
		\scalemath{1.3}{\Lambda(\phi)=[1, e^{-j2\pi\frac{d}{\lambda}\sin{\phi}},\ldots,e^{-j2\pi(\mathcal{V}-1)\frac{d}{\lambda}\sin{\phi}}]},  
	\end{align}\hrulefill
\end{figure*}

Similarly, the channel gains from BST to the $U^k_{n}$ and $U^k_{f}$, denoted as $\mathbf{h}^k_n$ and $\mathbf{h}^k_f$, respectively, being served in $k^{th}$ cluster are given as 
\begin{align}
	&\mathbf{h}^k_n=\sqrt{\eta_{0}\Big(\frac{d_{b,n}}{d_0}\Big)^{-\mu_{b,n}}}\overline{\mathbf{h}}^k_n, \label{7}
\end{align}
and

\begin{align}
	&\mathbf{h}^k_f=\sqrt{\eta_{0}\Big(\frac{d_{b,f}}{d_0}\Big)^{-\mu_{b,f}}}\overline{\mathbf{h}}^k_f, \label{8}
\end{align}    
where $d_{b,n}$ and $d_{b,f}$ denote the distance from BST to $U^k_{n}$ and $U^k_{f}$, respectively. $\mu_{b,n}$ and $\mu_{b,f}$ are the path-loss exponents from BST to $U^k_{n}$ and $U^k_{f}$, respectively. Moreover, $\overline{\mathbf{h}}^k_n$ and $\overline{\mathbf{h}}^k_f$ are considered under Rician fading transmission given as follows

 \begin{align}
	&	\overline{\mathbf{h}}^k_n = \sqrt{\frac{\delta_{1}}{1+\delta_{1}}}{{\mathbf{h}}^k_n}^{(LoS)}+\sqrt{\frac{1}{1+\delta_{1}}}{{\mathbf{h}}^k_n}^{(NLoS)}, \label{9}
\end{align}
and

\begin{align}
	&	\overline{\mathbf{h}}^k_f = \sqrt{\frac{\delta_{2}}{1+\delta_{2}}}{{\mathbf{h}}^k_f}^{(LoS)}+\sqrt{\frac{1}{1+\delta_{2}}}{{\mathbf{h}}^k_f}^{(NLoS)}, \label{10}
\end{align} 
where ${{\mathbf{h}}^k_n}^{(LoS)} \in \mathbb{C}^{N\times 1}$ and ${{\mathbf{h}}^k_f}^{(LoS)} \in \mathbb{C}^{N\times 1}$ denote the line-of-sight (LOS) components, whereas, $\delta_{1}$ and $\delta_{2}$ are the Rician factors for $U^k_{n}$ and $U^k_{f}$, respectively. Also, ${{\mathbf{h}}^k_n}^{(NLoS)} \in \mathbb{C}^{N\times 1}$, ${{\mathbf{h}}^k_f}^{(NLoS)} \in \mathbb{C}^{N\times 1}$ denote the non line-of-sight (NLOS) components which are i.i.d complex Gaussian random variables in nature with zero-mean and unit-variance. Moreover, the LOS components ${{\mathbf{H}}^k_n}^{(LoS)}$, ${{\mathbf{H}}^k_f}^{(LoS)} $, ${{\mathbf{h}}^k_n}^{(LoS)}$, and ${{\mathbf{h}}^k_f}^{(LoS)}$, from AP to BST and BST to IoT user in $k^{th}$ cluster, respectively, depend upon the ULA response at the AP and multi-antenna assisted BST given as follows 

\begin{align}
	&{{\mathbf{H}}^k_n}^{(LoS)} = \Lambda^H({\Phi^{arr.}_{tag}})\Lambda({\tilde{\Phi}^{dep.}_{AP}}), \label{11}
\end{align}
\begin{align}
	&{{\mathbf{H}}^k_f}^{(LoS)} = \Lambda^H({\Phi^{arr.}_{tag}})\Lambda({\tilde{\Phi}^{dep.}_{AP}}), \label{12}
\end{align}
and
\begin{align}
	&{{\mathbf{h}}^k_{n}}^{(LoS)} =\Lambda({\tilde{\Phi}^{dep.}_{tag}}), \label{13}
\end{align}
\begin{align}
	&{{\mathbf{h}}^k_{f}}^{(LoS)} =\Lambda({\tilde{\Phi}^{dep.}_{tag}}), \label{14}
\end{align}
where $\Lambda({\Phi^{arr.}_{tag}})$ represents the ULA response vector for angle of arrival at the multi-antenna assisted backscatter tag \cite{li2022robust}. Similarly, $\Lambda({\tilde{\Phi}^{dep.}_{AP}})$ is the ULA response vector for angle of departure at the AP node. $\Lambda({\tilde{\Phi}^{dep.}_{tag}})$ denotes the ULA response for angle of departure at BST. Furthermore, the ULA response of $\mathcal{V}$ antennas ULA is given by equation (7), where $\phi$ denotes the angle of arrival or angle of departure. Also, $\lambda$ and $d$ denote the wavelength of the carrier wave and spacing between radiating elements, respectively \cite{li2022robust}.
 
Next, let, $\mathbf{g}=[g_1,g_2,\ldots, g_n]^T \in \mathbb{C}^{N\times 1}$ is vector of diagonal entries of passive beamforming matrix $\mathbf{G}$. Then Eq. \eqref{1} and Eq. \eqref{2} are expressed as follows 

\begin{align}
	y^{k}_{n,1}=&({\mathbf {f}^k_{n}}^H+\mathbf{g}^H{\mathbf {B}^k_{n}})\mathbf{w}_k s_k m_k+(\mathbf {f}^k_{n}\nonumber\\
	&+{{\mathbf {f}^k_{n}}^H}\mathbf{g}^H)\sum\limits_{{\substack{l=1 \\ l\neq k}}}^K\mathbf{w}_l s_l m_l+n^k_{n,1}, \label{16}
\end{align}
and
\begin{align}
	y^{k}_{f,1}=&({\mathbf {f}^k_{f}}^H+\mathbf{g}^H{\mathbf {B}^k_{f}})\mathbf{w}_k s_k m_k+\nonumber\\
	&({\mathbf {f}^k_{f}}^H+{\mathbf {B}^k_{f}}\mathbf{g}^H)\sum\limits_{{\substack{l=1 \\ l\neq k}}}^K\mathbf{w}_l s_l m_l+n^k_{f,1}, \label{17}
\end{align}
where $\mathbf {B}^k_{n}=diag({\mathbf{h}^k_{n}}^H)\mathbf{H}^k_n$ and  $\mathbf {B}^k_{f}=diag({\mathbf{h}^k_{f}}^H)\mathbf{H}^k_f$ are the cascaded channel gains from AP to $U^k_{n}$ and $U^k_{f}$, respectively, through multi-antenna assisted BST.

Let, $\mathbf{v}^k_{n}={\mathbf {f}^k_{n}}^H+\mathbf{g}^H\mathbf {B}^k_{n}$ and $\mathbf{v}^k_{f}={\mathbf {f}^k_{f}}^H+\mathbf{g}^H\mathbf {B}^k_{f}$, then substituting the value of $s_k$ in Eq. \eqref{16} and Eq. \eqref{17}, we get Eq. (18) and Eq. (19), respectively.

Next, following the NOMA protocol, similar to the work in \cite{fang2017joint}, it is assumed that $||\mathbf{v}^k_{n}||^2 \geq||\mathbf{v}^k_{f}||^2$.  Hence, the desired signal-to-interference plus-noise ratio (SINR) at the $U^k_{n}$ node to determine the $U^k_{f}$ information in the $k^{th}$ cluster of the considered system can be expressed as follows
\begin{figure*}
\begin{align}
		\scalemath{1.1}
			{y^{k}_{n,1}=\underbrace{\mathbf{v}^k_{n}\mathbf{w}_k m_k(\sqrt{P_k\alpha_{k,n}}x_{k,n})}_\text{desired signal}}+
		\scalemath{1.1}{\underbrace{\mathbf{v}^k_{n}\mathbf{w}_k m_k(\sqrt{P_k\alpha_{k,f}}x_{k,f})}_\text{Intra-cluster interference}}\nonumber\\
			+\scalemath{1.1}{\underbrace{\sum\limits_{{\substack{l=1 \\ l\neq k}}}^K\mathbf{v}^k_{n}\mathbf{w}_l m_l( \sqrt{P_l\alpha_{l,n}}x_{l,n}+\sqrt{P_l \alpha_{l,f}}x_{l,f})}_\text{Inter-cluster Interference}}+
			\scalemath{1.1}{\underbrace{n^k_{n,1}}_\text{noise}},   
\end{align}\hrulefill
\end{figure*}

\begin{figure*}
	\begin{align}
		\scalemath{1}
		{y^{k}_{f,1}=\underbrace{\mathbf{v}^k_{f}\mathbf{w}_k m_k(\sqrt{P_k\alpha_{k,f}}x_{k,f})}_\text{desired signal}}+
		\scalemath{1}{\underbrace{\mathbf{v}^k_{f}\mathbf{w}_k m_k(\sqrt{P_k\alpha_{k,n}}x_{k,n})}_\text{Intra-cluster interference}}\nonumber\\
		+\scalemath{1}{\underbrace{\sum\limits_{{\substack{l=1 \\ l\neq k}}}^K\mathbf{v}^k_{f}\mathbf{w}_l m_l( \sqrt{P_l\alpha_{l,f}}x_{l,f}+\sqrt{P_l \alpha_{l,n}}x_{l,n})}_\text{Inter-cluster Interference}}+
		\scalemath{1}{\underbrace{n^k_{f,1}}_\text{noise}},   
	\end{align}\hrulefill
\end{figure*}
\begin{align}
	&	\gamma^k_{n,f}=\frac{P_{k} \alpha_{k,f}|\mathbf{v}^k_{n}\mathbf{w}_k|^2} {P_{k} \alpha_{k,n}| \mathbf{v}^k_{n}\mathbf{w}_k|^2+\Phi^k_{n,1}+\sigma^2}, \label{20}
\end{align}
where $\alpha_{k,n}$ denotes the PAC of $U^k_{n}$, $\Phi^k_{n,1}$ represents the inter-cluster interference for $k^{th}$ cluster can be expressed as 
\begin{align}
	&\Phi^k_{n,1}={\sum\limits_{{\substack{l=1 \\ l\neq k}}}^K|\mathbf{v}^k_{n}\mathbf{w}_l|^2 P_l(\alpha_{l,n}+\alpha_{l,f})}, \label{21}
\end{align}
Hence, the corresponding data rate at $U^k_{n}$ to decode $U^k_{f}$ information can be written as
\begin{align}
	R^k_{n,f} = \frac{1}{2}\log_{2}(1+\gamma^k_{n,f}), \label{22}
\end{align}

Similarly, the desired SINR and its corresponding data rate, after applying SIC, at the $U^k_{n}$ to decode its own data are expressed as follows
\begin{align}
	&	\gamma^k_{1}=\frac{P_{k} \alpha_{k,n}|\mathbf{v}^k_{n}\mathbf{w}_k|^2} {\Phi^k_{n,1}+\sigma^2}, \label{23}
\end{align}
and
\begin{align}
	R^k_{1} = \frac{1}{2}\log_{2}(1+\gamma^k_{1}), \label{24}
\end{align}

Further, the desired SINR at the $U^k_{f}$ to decode its information during the first time slot of considered BST-assisted IoT cooperative NOMA system can be written as follows
\begin{align}
	&\gamma^{k}_{2}=\frac{P_{k} \alpha_{k,f}|\mathbf{v}^k_{f}\mathbf{w}_k|^2} {P_{k} \alpha_{k,n}| \mathbf{v}^k_{f}\mathbf{w}_k|^2+\Phi^k_{f,1}+\sigma^2}, \label{25}
\end{align}
where $\Phi^k_{f,1}$ represents the ICI at the $U^k_{f}$ during first time frame and can be expressed as follows 
\begin{align}
	&\Phi^k_{f,1}={\sum\limits_{{\substack{l=1 \\ l\neq k}}}^K|\mathbf{v}^k_{f}\mathbf{w}_l|^2 P_l(\alpha_{l,f}+\alpha_{l,n})}, \label{26}
\end{align}
Hence, the corresponding rate for $\gamma^{k}_{2}$ is given as
\begin{align}
	R^{k}_{2} = \frac{1}{2}\log_{2}(1+\gamma^{k}_{2}), \label{27}
\end{align}
\subsection{Cooperative Transmission Phase}
In the $2^{nd}$ time frame of the considered BST-enabled cooperative IoT NOMA system, the signal received at $U^k_{f}$ in the $k^{th}$ cluster can be expressed as follows
\begin{align}
	y^{k}_{f,2}=\underbrace{\sqrt{P^k_{r}}h^k_{n,f} x^k_{n,f}}_\text{desired signal}
+\underbrace{\sum\limits_{{\substack{l=1 \\ l\neq k}}}^K\sqrt{P^l_r}h^{l(k)}_{n,f} x^l_{n,f}}_\text{Inter-cluster Interference}
+\underbrace{n^k_{f,2}}_\text{noise},\label{28}
\end{align}
where $P^k_{r}$ denotes the power of relay node in $k^{th}$ cluster, $h^k_{n,f}$ and $h^{l(k)}_{n,f}$ are complex Gaussian random variables, modeled as Rayleigh-fading, with zero-mean and unit-variance. $n^k_{f,2}$ represents the AWGN noise at $U^k_f$ with zero mean and variance $\sigma^2$. Hence, the desired SINR at the $U^k_{f}$ to decode its information during the $2^{nd}$ time slot of the considered system can be written as    
\begin{align}
	&\gamma^{k}_{3}=\frac{P^k_{r} |{h}^k_{n,f}|^2} {\Phi^k_{f,2}+\sigma^2}, \label{29}
\end{align}
where $\Phi^k_{f,2}$ is the ICI at the $U^k_{f}$ in second time slot and can be expressed as follows
\begin{align}
	&\Phi^k_{f,2}={\sum\limits_{{\substack{l=1 \\ l\neq k}}}^K P^l_r |h^{l(k)}_{n,f}|^2}, \label{30}
\end{align}

 As in \cite{maric2010bandwidth, laneman2004cooperative}, the resultant maximum rate achieved at $U^k_f$ can be written as:
\begin{align}
	R^k_{2}=min\Big\{\frac{1}{2}\Big(\log_{2}(1+\gamma^{k}_{n,f}), \log_{2}(1+(\gamma^{k}_{2}+\gamma^{k}_{3}))\Big)\Big\}, 
\end{align}
Based on decode-and-forward protocol, it is assumed that the source node successfully decodes the data for destination node. However, the $U^k_n$ can successfully decode the $U_f$ signal only if the rate of $U^k_f$ at $U^k_n$ is greater or equal to the maximum rate achieved at the $U^k_f$. Hence, to achieve successful cooperation, we consider that the following condition must be satisfied as as\cite{ali2019joint}:
\begin{align}
	&  \gamma^k_{nf}\geq \gamma^k_{2} + \gamma^k_{3}, \label{000000}
\end{align}

Consequently, an achievable sum-rate for $k^{th}$ cluster can be expressed as follows
\begin{align}
	R_{k} = R^k_{1}+ R^{k}_{2}, \label{32}
\end{align}
Next, the total power consumption for the $k^{th}$ cluster of considered BST-assisted cooperative IoT NOMA system can be written as
\begin{align}
	&P^k_{T}={\|\mathbf{w}_k\|^2 P_k(\alpha_{k,n}+\alpha_{k,f}) + P^k_{r}+P_c}, \label{33}
\end{align}
where $P_c$ denotes the circuit power consumption of the system. Herein, please note that the principal objective of this work is to maximize the energy efficiency (EE) of the considered BST-enabled multi-cluster cooperative IoT NOMA system by optimizing the power-allocation coefficients and active-beamforming at the transmitter and passive-beamforming at the multi-antenna assisted backscatter tag. Hence, under the desired QoS, SIC decoding, cooperation, power budget, and reflection-coefficient constraints, the proposed EE maximization problem (P) is formulated as  
\begin{subequations}\label{Prob:EE_1}
	\begin{align}
	\text{(P)}	\mathop {\max }\limits_{\boldsymbol{\alpha},\mathbf{w}_{k},\mathbf{g}}& EE=  \mathop {\max }\limits_{\boldsymbol{\alpha},\mathbf{w}_{k},\mathbf{g}}  \sum\limits_{k=1}^K  \frac{R_k}{P^k_{T}}\\
	s.t.\ & \gamma^k_{1}\geq \gamma^{\min}_{k,n},\forall k,\\
			\ & \gamma^{k}_{2}+\gamma^{k}_{3}\geq \gamma^{\min}_{k,f},\forall k,\\
			\ & \gamma^k_{nf}\geq \gamma^k_{2} + \gamma^k_{3},\forall k,\\
		&
		 \|\mathbf{w}_k\|^2 P_k\alpha_{k,f}|\mathbf{v}^k_{n}|^2-\|\mathbf{w}_k\|^2 P_k\alpha_{k,n}|\mathbf{v}^k_{n}|^2 \geq P_{gap} \\
		&\|\mathbf{w}_k\|^2 P_k (\alpha_{k,n}+\alpha_{k,f}) \leq P_{\max}, \forall k,\\
		& \alpha_{k,n}+\alpha_{k,f} \leq 1, \forall k,\\
		& 0 \leq {g}_{n} \leq 1, \forall n, 
	\end{align}
\end{subequations}
where $\boldsymbol{\alpha}=\{\alpha_{k,n},\alpha_{k,f}\}$, $\gamma^{\min}_{k,n}$, $\gamma^{\min}_{k,f}$ denotes the minimum required SINR to meet the QoS requirements at $U^k_{n}$ and $U^k_{f}$, respectively. Constraint (35d) guarantees the cooperation among the IoT NOMA users in the second time slot for the considered system, whereas, constraint (35e) ensures the successful SIC decoding, where $P_{gap}$ represents the minimum power difference desired to distinguish between the different IoT NOMA users for successful decoding \cite{ali2016dynamic}. Moreover, the Constraint defined in (35f) restricts the transmit power required to send the data for IoT NOMA users in each cluster based on the total available power budget $P_{\max}$. Finally, Constraints (35g) and (35h) limit power-allocation coefficients of the IoT NOMA users and reflection coefficients of multi-antenna BST within the practical range.     

Since, the objective function of the considered EE maximization problem defined in Eq. \eqref{Prob:EE_1} is non-convex for the coupled variables. Thus, it is very hard to tackle due to its non-linear fractional nature. Therefore, to reduce the complexity, successive convex approximation (SCA) \cite{papandriopoulos2009scale} can be exploited to transform the objective function into a tractable concave-convex fractional programming (CCFP) form. Thus, based on an SCA logarithmic approach, the sum-rate in \eqref{Prob:EE_1} can be expressed as follows:
\begin{align}
	\bar{R}_{k}=\sum_{i=1}^{2}\frac{1}{2}(\zeta_{i} \log_{2}(\gamma^k_{i})+\Gamma_{i}),  \label{35} 
\end{align} 
where $\zeta_{i}=\frac{\gamma^k_{i_o}}{1+\gamma^k_{i_o}}$ and $\Gamma_{i}=\log_{2}(1+\gamma^k_{i_o})-\frac{\gamma^k_{i_o}}{1+\gamma^k_{i_o}}\log_{2}(\gamma^k_{i_o})$ represent the approximation constants and the bound becomes tight at $\gamma^k_i=\gamma^k_{i_o}$.

Further, by exploiting the SCA approximation technique, the considered problem (P) is expressed as follows:
\begin{subequations}\label{Prob:EE_2}
	\begin{align}
		\text{(P1)}	\mathop {\max }\limits_{\boldsymbol{\alpha},\mathbf{w}_{k},\mathbf{g}}& EE=  \mathop {\max }\limits_{\boldsymbol{\alpha},\mathbf{w}_{k},\mathbf{g}}  \sum\limits_{k=1}^K  \frac{\bar{R}_{k}}{P^k_{T}}\\
		s.t.\	&  (34b) -(34h).
	\end{align}
\end{subequations}

\section{Solution for the Considered EE Maximization Problem for BST-enabled IoT Network}
The energy-efficiency maximization problem defined in Eq. \eqref{Prob:EE_2} is a non-convex function and it is very hard to solve for a global optimal solution due to coupled variables i.e., power allocation coefficients for IoT NOMA users $\alpha_{k,n}, \alpha_{k,f}$, active-beamforming vector $\textbf{{w}}_{k}$, and the passive-beamforming vector $\textbf{g}$ at the multi-antenna assisted backscatter tag. Thus, a sub-optimal solution is obtained for the considered EE maximization problem based on a proposed two-stage alternating optimization algorithm. i) In the $1^{st}$ stage, ZF-based active-beamforming and PAC $\alpha_{k,n}, \alpha_{k,f}$ are computed for the fixed value of the passive-beamforming vector $\g$ at the BST; ii) Subsequently, the passive-beamforming vector $\g$ is computed in the $2^{nd}$ stage for the PAC $\alpha_{k,n}, \alpha_{k,f}$ and active-beamforming vector $\textbf{{w}}_{k}$ obtained from stage one.

\subsection{ZF-Based Active-Beamforming and PAC: Stage 1}
The considered energy-efficiency maximization problem (P1) defined by Eq. \eqref{Prob:EE_2} can be simplified to an active-beamforming and power-allocation optimization problem for the given value of passive-beamforming vector $\g$ as follows 
\begin{subequations}\label{Prob:EE_3}
	\begin{align}
			\mathop {\max }\limits_{\boldsymbol{\alpha},\mathbf{w}_{k}}& EE=  \mathop {\max }\limits_{\boldsymbol{\alpha},\mathbf{w}_{k}}  \sum\limits_{k=1}^K  \frac{\bar{R}_{k}}{P^k_{T}}\\
		s.t.\ &  (34b) -(34g).
	\end{align}
\end{subequations}
The above EE maximization problem has a non-convex objective function coupled on two variables as $\boldsymbol{\alpha}$, $\textbf{w}_{k}$ and a set of non-convex constraints. Thus, to find an efficient solution for this non-convex problem, ZF-beamforming has been exploited to nullify the effect of ICI at the near $U^k_{n}$ in the $k^{th}$ cluster. Moreover, the NOMA-beamforming vectors are generated through the channel gains of strong IoT NOMA users in each cluster given as follows
\begin{align}
  	\mathbf{V}_k=[\mathbf{v}^1_{n},\ldots, \mathbf{v}^{k-1}_{n},\mathbf{v}^{k+1}_{n}\ldots,\mathbf{v}^{K}_{n}],\label{38} 
\end{align}
Hence, the ZF-beamforming constraint can be expressed as
\begin{align}
	&{\mathbf{V}_k}^H\mathbf{w}_k = 0, \forall k\label{39}.
\end{align}
where $\textbf{{w}}_{k}$ is the NOMA-beamforming vector for $k^{th}$ cluster as $\|{\mathbf{w}}_k\|^2 =1$.  Hence, the considered EE problem with ZF-formulation can be written as follows     
\begin{subequations}\label{Prob:EE_4}
	\begin{align}
		\mathop {\max }\limits_{\boldsymbol{\alpha},\mathbf{w}_{k}}& EE=  \mathop {\max }\limits_{\boldsymbol{\alpha},\mathbf{w}_{k}}  \sum\limits_{k=1}^K  \frac{\bar{R}_{k}}{P^k_{T}}\\
		s.t.\ &{\mathbf{V}_k}^H\mathbf{w}_k = 0, \forall k,\\
		&\|{\mathbf{w}}_k\|^2 =1, \forall k,\\
		&  (34b) -(34g). 
	\end{align}
\end{subequations}
Next, based on an efficient solution provided in \cite{shi2014joint}, optimal beamforming vector $\textbf{{w}}_{k}$ for the $k^{th}$ cluster can be computed as
\begin{align}
	&{\mathbf{w}}_k = \frac{\mathbf{Q}_k\mathbf{Q}_k^H\mathbf{v}^k_{n}}{\|\mathbf{Q}_k\mathbf{Q}_k^H\mathbf{v}^k_{n}\|^2} \label{41},
\end{align} 
where $\textbf{Q}_k$ represents the orthogonal-basis of the null-space of $\textbf{V}_k^H$ \cite{shi2014joint}. Note that an optimal active-beamforming vector $\mathbf{w}_k$, computed from Eq. \eqref{41}, not only increases the term $|\mathbf{v}^k_{n}\mathbf{w}_k|^2$ in the objective function, but also mitigates the effect of ICI $\Phi^k_{n,1}$ at the $U^k_{n}$ in each cluster of the considered BST-enabled cooperative IoT NOMA system. However, the weak IoT users $U^k_{f}$ in each cluster still receive the ICI. Thus, to minimize the effect of ICI at the $U^k_{f}$, an efficient clustering technique has been adopted based on the channel-gain difference and channel-correlation \cite{thet2021partial}. Specifically, only those IoT NOMA users are recruited for a cluster with strong channel-gain IoT NOMA users having maximum channel-correlation, as well as maximum channel gain difference with the strong IoT NOMA user. The expressions for channel-correlation and channel-gain difference for a pair of IoT NOMA users $(i, j)$ can be written as follows
\begin{align}
	\rho (i,j)= \frac{|\mathbf{v}_{i}^H\mathbf{v}_{j}|}{|\mathbf{v}_{i}^H||\mathbf{v}_{j}|}, \forall i,j\label{42}	
\end{align} 
\begin{align}
\delta(i,j)= ||\mathbf{v}_i||^2-||\mathbf{v}_j||^2, \forall i,j\label{42}	
\end{align}      
Hence, maximum correlation among clustered IoT users increases the term $|\mathbf{v}^k_{f}\mathbf{w}_k|^2$ which minimizes the effect of ICI $\Phi^k_{f,1}$ at the weak IoT user $U^k_{f}$ in $k^{th}$ cluster. Consequently, ZF-based active-beamforming, along with efficient clustering maximizes the EE of the considered BST-assisted cooperative IoT NOMA system.

Moreover, after the active-beamforming vector $\mathbf{w}_k$ and efficient clustering in hand, the optimization problem defined by Eq. \eqref{Prob:EE_4} can be written as
 
\begin{subequations}\label{Prob:EE_5}
	\begin{align}
		\mathop {\max }\limits_{\boldsymbol{\alpha}}& EE=  \mathop {\max }\limits_{\boldsymbol{\alpha}}  \sum\limits_{k=1}^K  \frac{\bar{R}_{k}}{P^k_{T}}\\
		s.t.\ & \gamma^k_{1}\geq \gamma^{\min}_{k,n},\forall k,\\
		\ & \gamma^{k}_{2}+\gamma^{k}_{3}\geq \gamma^{\min}_{k,f},\forall k,\\
		\ &  \gamma^k_{nf}\geq \gamma^k_{2} + \gamma^k_{3},\forall k,\\
		&
		\|\mathbf{w}_k\|^2 P_k\alpha_{k,f}|\mathbf{v}^k_{n}|^2-\|\mathbf{w}_k\|^2 P_k\alpha_{k,n}|\mathbf{v}^k_{n}|^2 \geq P_{gap} \\
		&\|\mathbf{w}_k\|^2 P_k (\alpha_{k,n}+\alpha_{k,f}) \leq P_{\max}, \forall k,\\
		& \alpha_{k,n}+\alpha_{k,f} \leq 1, \forall k, 
	\end{align}
\end{subequations}
Based on Dinkelbach algorithm \cite{dinkelbach1967nonlinear}, the above non-linear CCFP optimization problem in Eq. \eqref{Prob:EE_5} can be transformed to an equivalent parametric subtractive form given as \cite{fang2017joint}:
\begin{subequations}\label{Prob:EE_6}
	\begin{align}
	\mathop {\max }\limits_{\bm{\alpha}} &EE =\mathop  {\max }\limits_{\bm{\alpha}} \sum\limits_{k=1}^K  {\bar{R}_k}-\varrho_k P^k_{T}\\
	s.t.\ &  (44b) -(44g). 
\end{align}
\end{subequations}
where $\varrho_k$ represents the scaling parameter for $P^k_{T}$. Further, consider a function $\Upsilon(\varrho_k)$ given as follows 
\begin{align}
	\Upsilon(\varrho_k)=\underset{\bm{\alpha}}{\text{max}}  \{\bar{R}_{k}-\varrho_k P^k_{T}\},\label{46}
\end{align}
where $\Upsilon(\varrho_k)$ returns a negative quantity when $\varrho_k$ approaches to $\infty$ and positive quantity for $\varrho_k$ approaching to $-\infty$. Hence, $\Upsilon(\varrho_k)$ is affine with respect to $\varrho_k$. Specifically, solving the EE maximization problem in \eqref{Prob:EE_5} is analogous for computing the maximum-energy efficiency $\varrho^*_k$ \cite{fang2017joint}. Moreover, $\varrho^*_k$ can be obtained as long as the following condition is satisfied:
\begin{alignat}{2}
	\Upsilon(\varrho^*_k)=\underset{\bm{\alpha}}{\text{max}}  \{\bar{R}_{k}-\varrho^*_k P^k_{T}\}=0,\label{47}
\end{alignat}
 Next, Lagrange dual method and sub-gradient method have been exploited to find a sub-optimal solution for EE maximization problem defined in Eq. \eqref{Prob:EE_5}. Hence, the Lagrangian function of the optimization problem in \eqref{Prob:EE_5} can be formulated as follows
\begin{align} 
	&\mathcal{L}(\bm{\alpha},\bm{\varphi})= {\bar{R}_k}-\varrho_k P^k_{T}+\varphi_1\{\alpha_{k,n}\psi_n -(\bar{\sigma}^2_n\gamma^{\min}_{k,n}) \} \nonumber\\
	&+\varphi_2 \left[\alpha_{k,f}\psi_f -\left\{(\gamma^{\min}_{k,f}-\omega_1)\times(\alpha_{k,n}\psi_f+\bar{\sigma}^2_f)\right\}\right] \nonumber\\
		&+\varphi_{3}[\left(\alpha_{k,f}\psi_n \bar{\sigma}^{2}_{f}-\alpha_{k,f}\psi_{f} \bar{\sigma}^{2}_{n}\right)\nonumber\\ &-\left(\alpha^2_{k,n}\vartheta_{1}\omega_1+\alpha_{k,n}\vartheta_{2}\omega_1 + \bar{\sigma}^2_n\omega_1\right)] \nonumber\\
			&+\varphi_4 \left\{\|\mathbf{w}_k\|^2 P_k\alpha_{k,f}|\mathbf{v}^k_{n}|^2-\|\mathbf{w}_k\|^2 P_k\alpha_{k,n}|\mathbf{v}^k_{n}|^2 - P_{gap}\right\}\nonumber\\
	&+\varphi_5 \left\{P_{max}-\|\mathbf{w}_k \|^2  P_k(\alpha_{k,n}+\alpha_{k,f})\right\}\nonumber\\
		&+\varphi_6 \left\{1-(\alpha_{k,n}+\alpha_{k,f})\right\} \label{49}
\end{align}
where $\boldsymbol{\varphi}=\{\varphi_{1},\varphi_{2}, \varphi_{3}, \varphi_{4}, \varphi_{5}, \varphi_{6}\}$ are the Lagrange multipliers, $\psi_n=P_k|\mathbf{v}^k_{n}\mathbf{w}_k|^2$, $\psi_f=P_k|\mathbf{v}^k_{f}\mathbf{w}_k|^2$, $\bar{\sigma}^{2}_{n}=\Phi^k_{n,1}+\sigma^2$, $\bar{\sigma}^{2}_{f}=\Phi^k_{f,1}+\sigma^2$, $\omega_1=\gamma^{k}_{3}$, $\vartheta_{1}=\psi_n \psi_f$, $\vartheta_{2}=\psi_n \bar{\sigma}^{2}_{f}+ \psi_f \bar{\sigma}^{2}_{n}$ 

Further, Based on the Karush-Kuhn-Tucker (KKT) conditions, we can write as
\begin{alignat}{2}
\frac{\partial\mathcal{L}(\bm{\alpha},\bm{\varphi})}{\partial \bm{\alpha} }=0, \label{49} 
\end{alignat}
After taking the partial derivative of the Lagrangian function with respect to $\alpha_{k,n}$, we can write as 
\begin{align}
	& \frac {\zeta_{1}}{\alpha_{k,n}N}- \frac {\psi^2_f\zeta_2\alpha_{k,f}}{\alpha^2_{k,n}\delta_{1}+\alpha_{k,n}\delta_{1}+N_1}-(\alpha_{k,n}\Omega_1+\Omega_2)\nonumber\\
	&+\mu= 0 \label{50}
\end{align}
where $N=2\ln(2)$, $N_1=\omega_1\bar{\sigma}^{4}_{f}2\ln(2)$, $\delta_{1}=\omega_1\psi^2_f 2\ln(2)$, $\Omega_1=2\varphi_{3}\vartheta_{1}\omega_1$, $\Omega_2=\varphi_{3}\vartheta_{2}\omega_1$, $\delta_{2}=2\psi_{f}\omega_1\bar{\sigma}^{2}_{f}2\ln(2)+\psi^2_{f}\alpha_{k,f}2\ln(2)$.

Finally, after simplifying the Eq. \eqref{50}, we obtain a quartic polynomial equation which can be easily solved by employing the built-in functions provided in MATLAB and Mathematica solvers to compute optimal power-allocation coefficients $\alpha^*_{k,n}$, $\alpha^*_{k,f}$ as follows
\begin{align}
\alpha^*_{k,n}=\Big(\alpha^4_{k,n}\tilde{\Psi}_{4}+\alpha^3_{k,n}\tilde{\Psi}_{3}+\alpha^2_{k,n}\tilde{\Psi}_{2}+\alpha_{k,n}\tilde{\Psi}_{1}+ \tilde{\Psi}=0\Big) ,\label{51}  
\end{align}
and, \begin{align}
	\alpha^*_{k,f}=1-\alpha^*_{k,n},\label{511}  
\end{align}
where $\tilde{\Psi}_{4}=-N\Omega_1\delta_1$, $\tilde{\Psi}_{3}=(N\mu\delta_1-N\Omega_2\delta_1-N\Omega_1\delta_{2})$,  $\tilde{\Psi}=N_1\zeta_{1}$. The values of $\tilde{\Psi}_{1}$, $\tilde{\Psi}_{2}$ and $\mu$ are given in Eq. (54) and Eq. (55), and Eq. (56) respectively. 
\begin{figure*}
	\begin{align}
	\scalemath{1.1}{\tilde{\Psi}_{1} = \zeta_{1}\delta_{2}-N\psi^2_{f}\alpha_{k,f}\zeta_{2}- NN_1\Omega_2+ NN_1\mu},     
	\end{align}\hrulefill
\end{figure*}

\begin{figure*}
	\begin{align}
		\scalemath{1.1}{\tilde{\Psi}_{2} = \zeta_{1}\delta_{1}-NN_1\Omega_1- N\Omega_2\delta_{2}+ N\delta_{2}\mu},     
	\end{align}\hrulefill
\end{figure*}

\begin{figure*}
\begin{align}
	\scalemath{1.1}{\mu= \varphi_{1}\psi_n-\varrho P_k\|\mathbf{w}_k\|^2 - \varphi_{2}\psi_{f}(\gamma^{\min}_{k,f}-\omega_1)-\varphi_{4}\psi_n - \varphi_{5}P_k\|\mathbf{w}_k\|^2-\varphi_{6}},     
\end{align}\hrulefill
\end{figure*}   

Subsequently, for a given power-allocation policy for the IoT NOMA users in Eq. \eqref{51} and Eq. \eqref{511}, the Lagrange multipliers $\varphi_{1}$, $\varphi_{2}$, $\varphi_{3}$, $\varphi_{4}$, $\varphi_{5}$, and $\varphi_{6}$ are iteratively updated based on sub-gradient method as \cite{khan2019joint}:
\begin{align}
	\varphi_{1}(i+1)=\left[\varphi_{1}(i)+\Upsilon_1(i)   
	\Big(\alpha_{k,n}\psi_n -\bar{\sigma}^2_n\gamma^{\min}_{k,n}\Big)\right]^+ ,\label{55}  
\end{align}

\begin{align}
	\varphi_{2}(i+1)=\left[\varphi_{2}(i)+\Upsilon_2(i)   
	\Big(\alpha_{k,f}\psi_f -\gamma^{\min}_{k,f}(\alpha_{k,n}\psi_f+\bar{\sigma}^2_f)\Big)\right]^+ ,\label{56}  
\end{align}
\begin{align}
	\varphi_{3}(i+1)=&\Big[\varphi_{3}(i)+\Upsilon_3(i)   
	\Big\{(-\alpha^2_{k,n}(\omega_1\psi_n\psi_f)-\alpha_{k,n}(\omega_1\bar{\sigma}^2_n\psi_f\nonumber\\
	&+\omega_1\psi_n\bar{\sigma}^2_f)+ (\alpha_{k,f}\psi_n\bar{\sigma}^2_f-\bar{\sigma}^2_n\alpha_{k,f}\psi_f\nonumber\\
	&-\bar{\sigma}^2_n\bar{\sigma}^2_f\omega_1)\Big\}\Big]^+ ,\label{57}  
\end{align}
\begin{align}
	\varphi_{4}(i+1)=&\Big[\varphi_{4}(i)+\Upsilon_4(i)   
	\Big\{(\|\mathbf{w}_k\|^2 P_k\alpha_{k,f}|\mathbf{v}^k_{n}|^2) \nonumber\\
	&-(\|\mathbf{w}_k\|^2 P_k\alpha_{k,n}|\mathbf{v}^k_{n}|^2 - P_{gap})\Big\}\Big]^+ ,\label{58}  
\end{align}
\begin{align}
	\varphi_{5}(i+1)=&\Big[\varphi_{5}(i)+\Upsilon_5(i)   
	\Big\{P_{max}-\|\mathbf{w}_k \|^2  P_k(\alpha_{k,n}+\alpha_{k,f})\Big\}\Big]^+ ,\label{59}  
\end{align}

\begin{align}
	\varphi_{6}(i+1)=&\Big[\varphi_{6}(i)+\Upsilon_6(i)   
	\Big\{1-(\alpha_{k,n}+\alpha_{k,f})\Big\}\Big]^+ ,\label{60}  
\end{align}
where $\Upsilon_1(i)$, $\Upsilon_2(i)$, $\Upsilon_3(i)$, $\Upsilon_4(i)$, $\Upsilon_5(i)$, and $\Upsilon_6(i)$ are the step sizes and $i$ represents the iteration index. However, appropriate step sizes are very crucial to get the convergence of algorithm in an iterative manner.

\subsection{Passive-Beamforming Optimization: Stage 2}
For the optimal values of active-beamforming vector and PAC of IoT NOMA users obtained in Stage 1, the considered EE maximization optimization problem for passive-beamforming vector $\textbf{g}$ can be expressed as follows
\begin{subequations}\label{Prob:EE_7}
	\begin{align}
		\mathop {\max }\limits_{\mathbf{g}}& EE=  \mathop {\max }\limits_{\mathbf{g}}  \sum\limits_{k=1}^K {\bar{R}_k}-\varrho_k P^k_{T}\\
		s.t.\ & \gamma^k_{1}\geq \gamma^{\min}_{k,n},\forall k,\\
		\ & \gamma^{k}_{2}+\gamma^{k}_{3}\geq \gamma^{\min}_{k,f},\forall k,\\
		\ &  \gamma^k_{nf}\geq \gamma^k_{2} + \gamma^k_{3},\forall k,\\
		&
		\|\mathbf{w}_k\|^2 P_k\alpha_{k,f}|\mathbf{v}^k_{n}|^2-\|\mathbf{w}_k\|^2 P_k\alpha_{k,n}|\mathbf{v}^k_{n}|^2 \nonumber\\
		&\geq P_{gap} \\
 	    & 0 \leq {g}_{n} \leq 1, \forall n,
	\end{align}
\end{subequations}

Further, the main steps to compute the optimal passive-beamforming vector $\mathbf{g}$ are summarized as follows:

\textbf{Step 1:} Let, $\textbf{E}^k_n=\textbf{B}^k_n {\textbf{w}_{k}}$,  $\textbf{E}^k_f=\textbf{B}^k_f {\textbf{w}_{k}}$, ${\textbf{f}^k_{n}}^H {\textbf{w}_{k}}=\tau^k_n$, ${\textbf{f}^k_{f}}^H {\textbf{w}_{k}}=\tau^k_f$, and $\mathbf{\bar{g}}={[\g ; 1]}^H$. Further, the sum-rate for the optimization problem in \eqref{Prob:EE_7} can be written as
\begin{align}
	\bar{R}_{k} =\bar{R}^k_{1}+ \bar{R}^{k}_{2}, \label{62}
\end{align}  
Let us define a matrix $\mathbf{F}=\mathbf{\bar{g}}\mathbf{\bar{g}}^H$, where $\mathbf{F} \succeq 0$, $rank (\mathbf{F}) = 1$. Then,
\begin{align}
	\bar{R}^k_{1} &=\Big[\zeta_{1}\Big\{\log_{2}\Big(P_k\alpha_{k,n} \times \Big(Tr.(\mathbf{M}^k_n\mathbf{F})+|\tau^k_n|^2\Big)\Big)\nonumber\\
	&-\log_{2}(\bar{\sigma}^{2}_{n})\Big\} +\Gamma_{1}\Big], \label{63}
\end{align} 
and,
\begin{align}
	\bar{R}^k_{2} &=\Big[\zeta_{2}\Big\{\log_{2}\Big(P_k\alpha_{k,f} \times \Big(Tr.(\mathbf{M}^k_f\mathbf{F})+|\tau^k_f|^2\Big)+\nonumber\\
	&\omega_1(P_k\alpha_{k,n} \Big(Tr.(\mathbf{M}^k_f\mathbf{F})+|\tau^k_f|^2\Big)+\bar{\sigma}^{2}_{f})\Big)\nonumber\\
	&-\log_{2}\Big(P_k\alpha_{k,n} \Big(Tr.(\mathbf{M}^k_f\mathbf{F})+|\tau^k_f|^2\Big)+\bar{\sigma}^{2}_{f}\Big)\Big\} +\Gamma_{2}\Big], \label{64}
\end{align}
where $Tr.(.)$ denotes the trace function, and
\begin{align}
\mathbf{M}^k_{n}=\begin{bmatrix}
\textbf{E}^k_n {\textbf{E}^k_n}^H	& \textbf{E}^k_n {\tau^k_n}^H\\ 
 {\textbf{E}^k_n}^H \tau^k_n& 0
\end{bmatrix},\label{66}
\end{align}
\begin{align}
	\mathbf{M}^k_{f}=\begin{bmatrix}
		\textbf{E}^k_f {\textbf{E}^k_f}^H	& \textbf{E}^k_f {\tau^k_f}^H\\ 
		{\textbf{E}^k_f}^H \tau^k_f& 0
	\end{bmatrix},\label{67}
\end{align}
   
\textbf{Step 2:} Since $\bar{R}^k_{2}$ in the Eq. \eqref{64} consists of the difference of two functions concave in nature. Hence, the EE maximization problem given in \eqref{Prob:EE_7} is non-convex. We can write $\bar{R}^k_{2}$ as a function of $\textbf{F}$ given as 
\begin{align}
	\bar{R}^{k*}_{2} &=\Big[\zeta_{2}\Big(\Psi_1(\mathbf{F})-\Psi_2(\mathbf{F})\Big) +\Gamma_{2}\Big], \label{68}
\end{align}         
Further, we exploit a low-complexity sub-optimal technique, known as difference of convex (DC) programming, to transform the above non-convex optimization problem into a convex EE maximization problem \cite{shen2018fractional}. Based on the DC programming technique, instead of having $\Psi_2(\textbf{F})$, we use its first-order Taylor expansion as  
\begin{align}
	\Psi_2(\mathbf{F}) & \leq \Psi_2(\mathbf{F}^{(m)}+Tr.\Big(\big(\Psi_2^{'}(\mathbf{F}^{(m)})\big)^{H}\big(\mathbf{F}-\mathbf{F}^{(m)}\big)\Big), \nonumber\\
	& \triangleq \overline{\Psi_2(\mathbf{F})} \label{69}
\end{align}
where $\mathbf{F}^{(m)}$ is the value of $\mathbf{F}$ in the $m^{th}$ iteration and $\Psi_2^{'}(\mathbf{F}^{(m)})$ represents the first-order derivative of $\Psi_2(\mathbf{F})$ in $m^{th}$ iteration. 

Based on the first-order Taylor expansion of $\Psi_2(\textbf{F})$, we can write as  
\begin{align}
	\bar{R}^{k**}_{2} &=\Big[\zeta_{2}\Big(\Psi_1(\mathbf{F})- \overline{\Psi_2(\mathbf{F})}\Big) +\Gamma_{2}\Big], \label{70}
\end{align}

 Next, based on Eq. \eqref{70}, the sum-rate can be updated as  
\begin{align}
	\tilde{R}_{k} =\bar{R}^k_{1}+ \bar{R}^{k**}_{2}, \label{71}
\end{align}
Hence, the optimization problem in Eq. \eqref{Prob:EE_7} can be reformulated as follows 
\begin{subequations}\label{Prob:EE_8}
	\begin{align}
		\mathop {\max }\limits_{\mathbf{F}}& EE=  \mathop {\max }\limits_{\mathbf{F}}  \sum\limits_{k=1}^K {\tilde{R}_k}-\varrho_k P^k_{T}\\
		s.t.\ & P_k\alpha_{k,n}\Big(Tr.(\mathbf{M}^k_n\mathbf{F})+|\tau^k_n|^2\Big)\geq \bar{\sigma}^{2}_{n}\gamma^{\min}_{k,n},\forall k,\\
		       \ & P_k\alpha_{k,f}\Big(Tr.(\mathbf{M}^k_f\mathbf{F})+|\tau^k_f|^2\Big) \geq (\gamma^{\min}_{k,f}-\omega_1) \nonumber\\
		&\times (P_k\alpha_{k,n} \Big(Tr.(\mathbf{M}^k_f\mathbf{F})+|\tau^k_f|^2\Big)+\bar{\sigma}^{2}_{f}),\forall k,\\
		           \ & \Big( P_k\alpha_{k,f}\bar{\sigma}^{2}_{f}\Big(Tr.(\mathbf{M}^k_n\mathbf{F})+|\tau^k_n|^2\Big)\Big)-\Big( P_k\alpha_{k,f}\bar{\sigma}^{2}_{n}\nonumber\\
		           &\Big(Tr.(\mathbf{M}^k_f\mathbf{F})+|\tau^k_f|^2\Big)\Big) \geq \gamma^{k}_{3}\Big[\Big(P_k\alpha_{k,n}\Big(Tr.(\mathbf{M}^k_n\mathbf{F})\nonumber\\
		           &+|\tau^k_n|^2\Big)+ \bar{\sigma}^{2}_{n} \Big)\times\Big(P_k\alpha_{k,n}\Big(Tr.(\mathbf{M}^k_f\mathbf{F})+|\tau^k_f|^2\Big)+ \bar{\sigma}^{2}_{f} \Big)\Big] \\
		                   & P_k\alpha_{k,f}\Big(Tr.(\mathbf{M}^k_n\mathbf{F})+|\tau^k_n|^2\Big)- P_k\alpha_{k,n}\Big(Tr.(\mathbf{M}^k_n\mathbf{F})\nonumber\\
		                   &+|\tau^k_n|^2\Big) \geq P_{gap} \\
		                         & \mathbf{F}_{n,n} \leq 1, \forall n,\\
		                            &\mathbf{F} \succeq 0,\\
		                               &rank (\mathbf{F}) = 1. 
	\end{align}
\end{subequations}

\textbf{Step 3:} The above EE maximization optimization problem is still non-convex in nature because of the Rank-1 constraint defined in Eq. (73h). Since $\mathbf{F} \in \mathbb{C}^{N\times N}$, Tr.($\mathbf{F})>0$, is positive semi definite (PSD) matrix, we exploit DC programming to transform this Rank-1 constraint as follows
\begin{equation}
	rank (\mathbf{F}) = 1 \Leftrightarrow Tr.(\mathbf{F}) -||\mathbf{F}||_2 =0, \label{73}
\end{equation}
where $Tr.(\mathbf{F}) = \sum\limits_{n=1}^{N} \rho_n(\mathbf{F})$ and $||\mathbf{F}||_2=\rho_1(\mathbf{F})$ denote the trace and spectral-norm of PSD matrix $\mathbf{F}$, respectively. $\rho_n$ represents the $n^{th}$ largest singular value of matrix $\mathbf{F}$. Since, Rank-1 constraint in Eq. \eqref{73} is still non-convex. Therefore, Rank-1 constraint has been transformed based on the Taylor expansion of spectral norm $||\mathbf{F}||_2$ as follows 
\begin{align}
	||\mathbf{F}||_2 &\geq ||\mathbf{F}^{(m)}||_2 Tr.\Big(\mathcal{E}^{(m)}_{max}(\mathbf{F}^{(m)}) {\mathcal{E}^{(m)}_{max}(\mathbf{F}^{(m)})}^H(\mathbf{F}-\mathbf{F}^{(m)})\Big)\nonumber , \\
	&  \triangleq \overline{||\mathbf{F}||_2},\label{74}
\end{align}
where $\mathcal{E}^{(m)}(\mathbf{F}^{(m)})$ denotes the eigenvector corresponding to the largest singular value of PSD matrix $\mathbf{F}$ in the $m^{th}$ iteration. It is worth noting that herein we choose \eqref{73} as one of the ways to solve the Rank-$1$ constraint, while we can also adopt some other practical methods that deal with the Rank-$1$ issue such as the unified manner by the semi-definite relaxation (SDR) technique \cite{luo2010semidefinite} or the  unified manner by the quadratically constrained quadratic problems (QCQP) forms with the Feasible Point Pursuit (FPP) algorithm \cite{christopoulos2015multicast}. It has been shown in \cite{xu2020resource} that the equivalence in \eqref{73} can already solve the Rank-$1$ problem fairly well.  

  \begin{algorithm}[!t]
	{\bf Initialization:} Initialize all of the system parameters  \\
	\While{not converge}{
		\textbf{Stage 1:} Compute $\mathbf{w}_k$ and PAC $\alpha_{k,n}$, $\alpha_{k,f}$, for the fixed value of passive-beamforming vector $\mathbf{g}$.\\
		\While{not converge}{
			\For{$k=1:K$}{Compute active-beamforming vector $\mathbf{w}_k$ usin Eq. \eqref{41}\\ 
				Compute PAC for the IoT NOMA user $U^k_n$ in the $k^{th}$ cluster, $\alpha_{k,n}$ using Eq. \eqref{51}\\
				Compute PAC for the IoT NOMA user $U^k_f$ in the $k^{th}$ cluster, $\alpha_{k,f}$ using Eq. \eqref{511}\\
				Update the Lagrange multipliers using Eqs. \eqref{55}--\eqref{60} }
		}
		\textbf{Stage 2:} With optimal $\mathbf{w}^*_k$, $\alpha^*_{k,n}$, and $\alpha^*_{k,f}$ in hand, compute $\mathbf{g}$\\
		\While{not converge}{
			\For{$k=1:K$}{
				\For{$n=1:N$}{Compute the optimal passive-beamforming vector $\mathbf{g}$ by solving the standard SDP optimization problem in Eq. \eqref{Prob:EE_9} using MOSEK-enabled CVX convex optimization toolbox for MATLAB.
			}}
		}
	}
	Return $\mathbf{w}^*_k$, $\alpha^*_{k,n}$, $\alpha^*_{k,f}$, $\mathbf{g}^*$
	\caption{Proposed Energy-Efficient Optimization Algorithm for BST-Assisted Cooperative IoT NOMA System}
\end{algorithm}  
 \begin{table*}[!t]
 	\captionsetup{justification=centering}
 	\caption{Simulation Parameters and Values}
 	\label{TABLE 1}
 	\centering
 	\begin{tabular}{| c | c |}
 		\hline
 		{\textbf{Parameters}} & {\textbf{Value}} \\
 		\hline
 		path-loss exponent ($\mu_{s,b}$ , $\mu_{b,n}$, $\mu_{b,f}$) & 2.2\\
 		\hline
 		Bandwidth & 1 MHz \\
 		\hline
 		Distance between AP and BST & 30m \\
 		\hline
 		Total number of clusters & 5 \\
 		\hline
 		Number of IoT NOMA users/cluster  & 2 \\
 		\hline
 		Number of active-beamforming vectors  & 5 \\
 		\hline
 		Noise-power ($\sigma^2$) & -114 dBm \\
 		\hline
 		Transmit power/cluster ($P_{k}$) &  30 dBm \\
 		\hline
 		Minimum SINR for QoS ($\gamma^{\min}_{k,n}, \gamma^{\min}_{k,f}$) & 3 dB \\\hline
 		Relay power/cluster & 10 dBm \\
 		\hline
 		Circuit power ($P_{c}$) & 0.1 W \\
 		\hline
 		BST radius & 10m\\
 		\hline
 		Fast-fading & Rayleigh-fading, Rician-fading \\
 		\hline
 		Rician factor ($\beta_{1},\beta_{2},\delta_{1},\delta_{2}$) & 3 \\
 		\hline
 		correlation threshold & $0$ -- $1$ \\
 		\hline
 		IoT NOMA users distribution model & Binomial-point process (BPP) \\
 		\hline
 	\end{tabular}
 \end{table*}
\textbf{Step 4:} Finally, we add the above transformed Rank-1 constraint into the objective function of the considered optimization problem defined in \eqref{Prob:EE_8} as a penalty term as follows
\begin{subequations}\label{Prob:EE_9}
	\begin{align}
		\mathop {\max }\limits_{\mathbf{F}}& EE=  \mathop {\max }\limits_{\mathbf{F}}  \sum\limits_{k=1}^K {\tilde{R}_k}-\varrho_k P^k_{T}-\xi\Big(Tr.(\mathbf{F})-\overline{||\mathbf{F}||_2}\Big)\\
	s.t.\ &  (73b) -(73g).
	\end{align}
\end{subequations}
where $\xi>>0$ denotes the penalty factor for the Rank-1 constraint. Consequently, the above EE maximization problem for the considered system, defined in Eq. \eqref{Prob:EE_9}, is a standard Semi-definite programming (SDP) optimization problem, where the solution of this SDP problem can be obtained by efficiently exploiting the CVX-enabled optimization toolkit integrated with MATLAB \cite{grant2014cvx}.

\subsection{Proposed Algorithm and Computational Complexity}

An energy-efficient alternating optimization algorithm for multi-antenna BST-assisted cooperative NOMA has been proposed. The proposed algorithm tackles the considered EE maximization problem in two stages. In the $1^{st}$ stage, active-beamforming vector $\mathbf{w}_k$ and PAC $\alpha_{k,n}$, $\alpha_{k,f}$ are computed for the fixed value of passive-beamforming vector $\g$. Once the $\mathbf{w}_k$ and PAC $\alpha_{k,n}$, $\alpha_{k,f}$ are in hand, passive-beamforming vector $\g$ is computed in the $2^{nd}$ stage of proposed \textbf{Algorithm 1}.   

The complexity of stage 1 for the considered alternation optimization algorithm is $O(K)$. In the $2^{nd}$ stage, the relaxed SDP problem is solved by exploiting the interior-point method with computational complexity $O(KN^{3.5})$ \cite{wright1997primal}. If $m$ represents the number of iterations required for the convergence of the proposed two-stage alternation optimization framework, then the total computational complexity of \textbf{Algorithm 1} can be expressed as $O\Big(m(K+KN^{3.5})\Big)$.

   \begin{figure}[!t]
   	\centering
   	\includegraphics [width=0.48\textwidth]{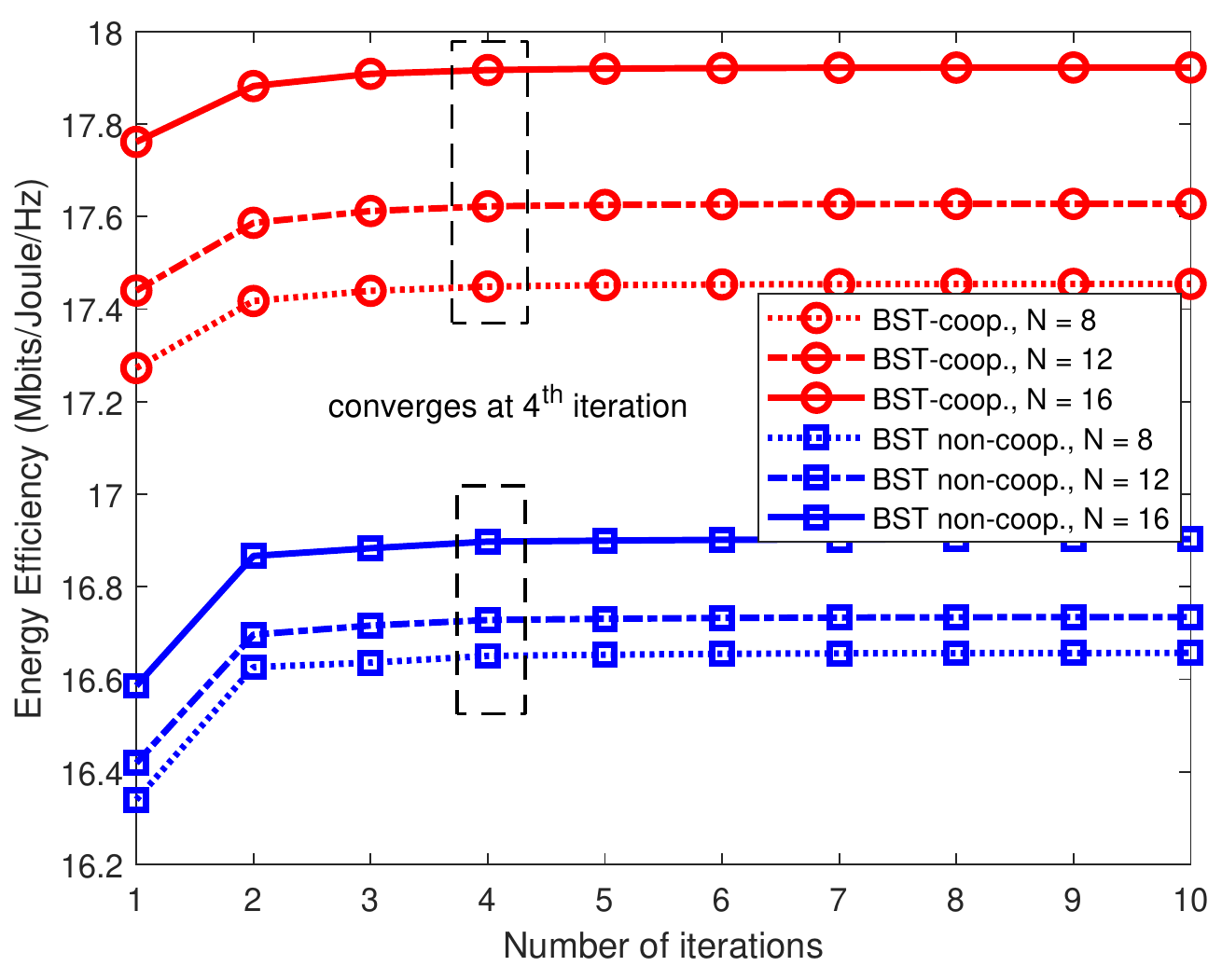}
   	\caption{The convergence of proposed BST-coop. NOMA and its competitor under different number of iterations}
   	\label{f2}
   \end{figure}

 \begin{figure}[!t]
	\centering
	\includegraphics [width=0.48\textwidth]{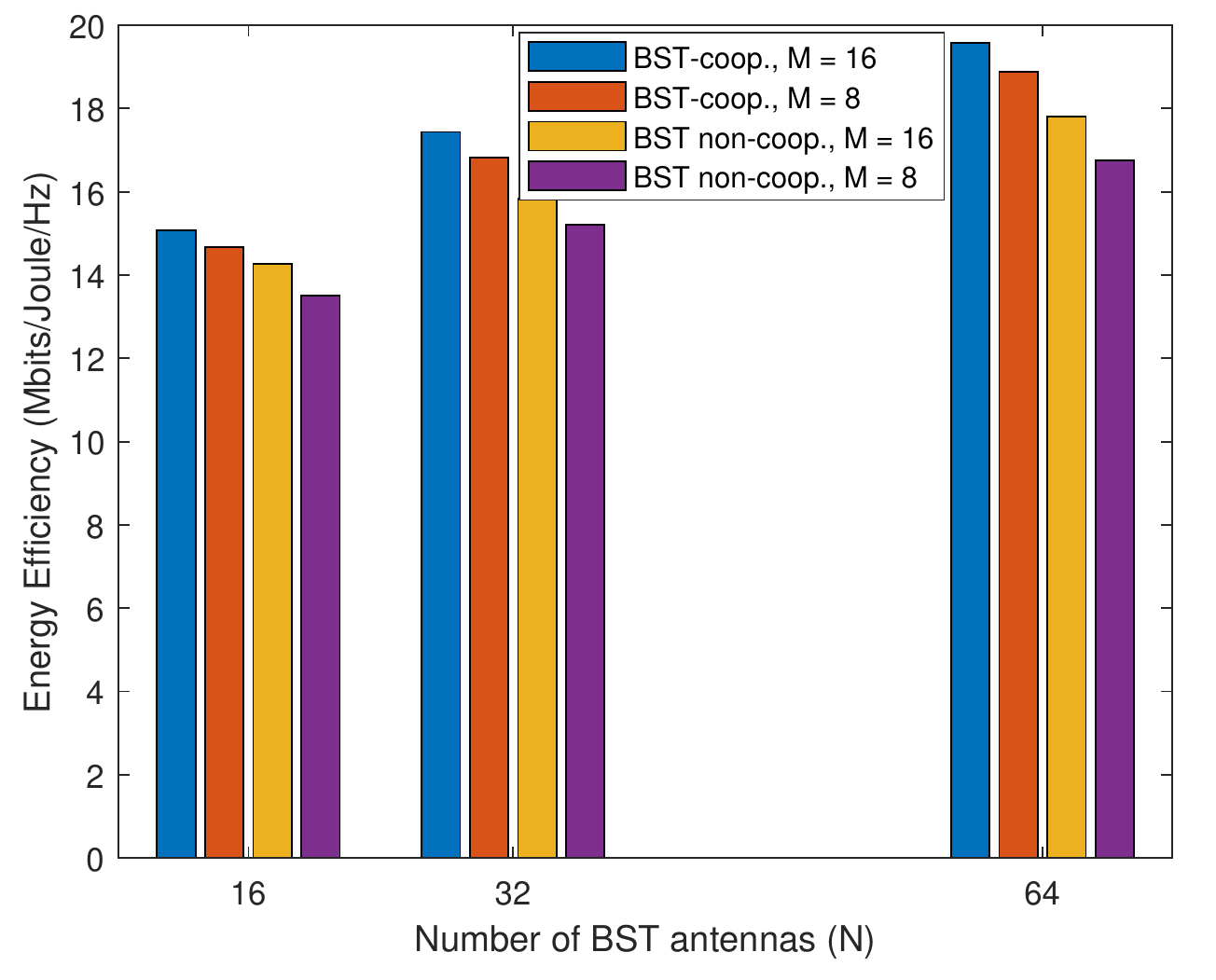}
	\caption{Th Energy-efficiency vs. different number of antennas at backscatter tag }
	\label{f3}
\end{figure}

\section{Numerical Simulation Results and Discussion}

This section presents the performance analysis, in terms of the energy-efficiency, for the considered multi-antenna BST-enabled cooperative IoT NOMA network. A comparative analysis, in terms of energy-efficiency, has been conducted to demonstrate the performance proposed BST-enabled cooperative IoT NOMA system and its non-cooperative NOMA counterpart, labeled as BST-coop. and BST non-coop., respectively, under various performance parameters. Further, Monte Carlo simulations, under $10^4$ channel realizations, have been exploited to obtain the simulation results, where 5 clusters are designed based on the considered clustering technique. A total of 30 IoT users are randomly deployed in a circle around the BST with a radius of 10m. The distance between AP and BST is considered as 30m, however, the positions of the randomly deployed IoT NOMA users around the BST are modeled as a binomial-point process. Also, $\eta_{0}=-30$ dB with $d_0=1$(m). Unless mentioned, the simulation parameters utilized in this work are given in TABLE \ref{TABLE 1}.

The convergence behavior of the proposed multi-antenna BST-assisted cooperative IoT NOMA system is depicted in  Fig. \ref{f2}, where the performance of the proposed BST coop. system is compared with its BST non-coop counterpart in terms of energy-efficiency for a different number of antennas realized at the BST. Fig. \ref{f2} evinces that only a few iterations are required for the convergence of the proposed algorithm. Also, it can be seen that the proposed BST coop. system provides an efficient energy-efficiency performance and outperforms its BST non-coop. competitor by for the different number of BST antennas. Furthermore, it is important to note that the energy-efficiency increases by increasing the BST antennas, however, it does not affect the convergence of the system. Moreover, It is important to mention that in most of the scenarios for the considered system, Rank-1 solution is obtained for passive-beamforming vector based on the penalty-based method which means that the equivalence in \eqref{73} can efficiently solve the Rank-1 problem in our scenario setting.
  
 \begin{figure}[!t]
	\centering
	\includegraphics [width=0.52\textwidth]{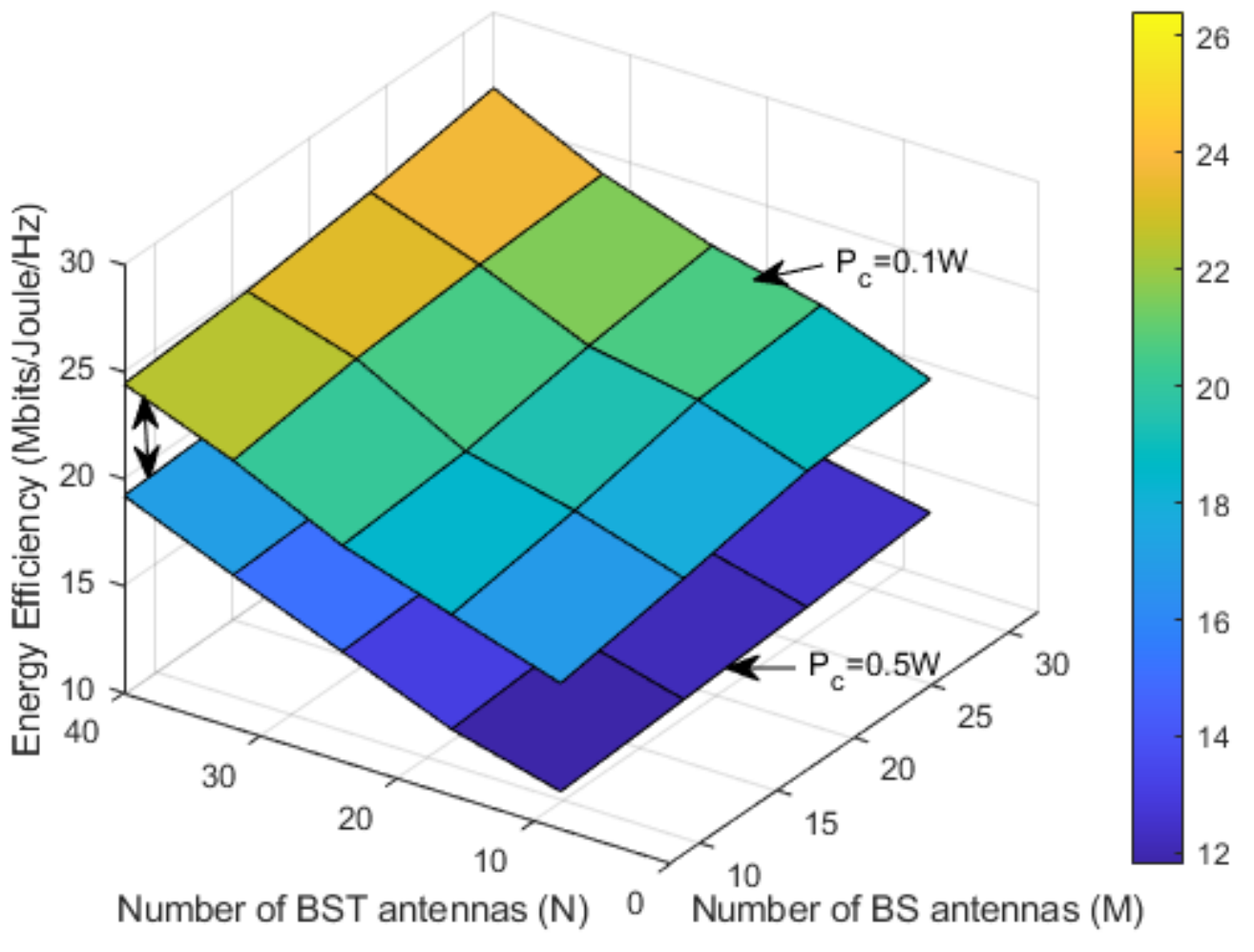}
	\caption{Energy-efficiency against different number of AP antennas $M$ and BST antennas $N$}
	\label{f4}
\end{figure} 

Next, the impact of a different number of antennas equipped at the backscatter node on the energy-efficiency of the system has been depicted in  Fig. \ref{f3}. It can be observed that the energy-efficiency of the system increases by increasing the number of BST antennas. The performance of the proposed BST coop. system is compared with its BST non-coop. counterpart under a different number of antennas at the AP. Simulation results in Fig. \ref{f3} evince that the performance of the proposed BST coop. system is superior over its BST non-coop. competitor in terms of energy-efficiency of the considered system.  

Fig. \ref{f4} reveals the impact of AP antennas $M$ and BST antennas $N$ on the energy-efficiency of the considered system. It can be observed that the energy-efficiency is more sensitive to BST antennas N as compared to AP antennas. For instance if we change AP antennas $M$ from $8$ to $32$ while keeping BST antennas constant with $N=8$ and $P_c=0.1W$, the energy-efficiency values are obtained as $17.82$  Mbits/Joule/Hz, $18.89$  Mbits/Joule/Hz, and $19.23$ Mbits/Joule/Hz, respectively. However, if BST antennas $N$ are varied from $8$ to $32$ with AP antennas $M=8$, the energy-efficiency is obtained as $18.46$  Mbits/Joule/Hz, $20.09$  Mbits/Joule/Hz, and, $22.46$ Mbits/Joule/Hz, respectively. This is because on increasing $M$ while keeping $N$ unchanged, the energy-efficiency of the system increases as the active-beamforming vector $\textbf{{w}}_{k}$ will have an approximately constant impact on energy-efficiency due to maximum power-budget constraint in (36f). However, increasing $N$ will have a prominent effect to increase the energy-efficiency through passive-beamforming by optimizing the reflection coefficients of the multi-antenna backscatter tag. Moreover, simulation results evince that higher values of energy-efficiency are achieved for lower value of circuit-power. 

\begin{figure}[t]
 	\centering
 	\includegraphics [width=0.48\textwidth]{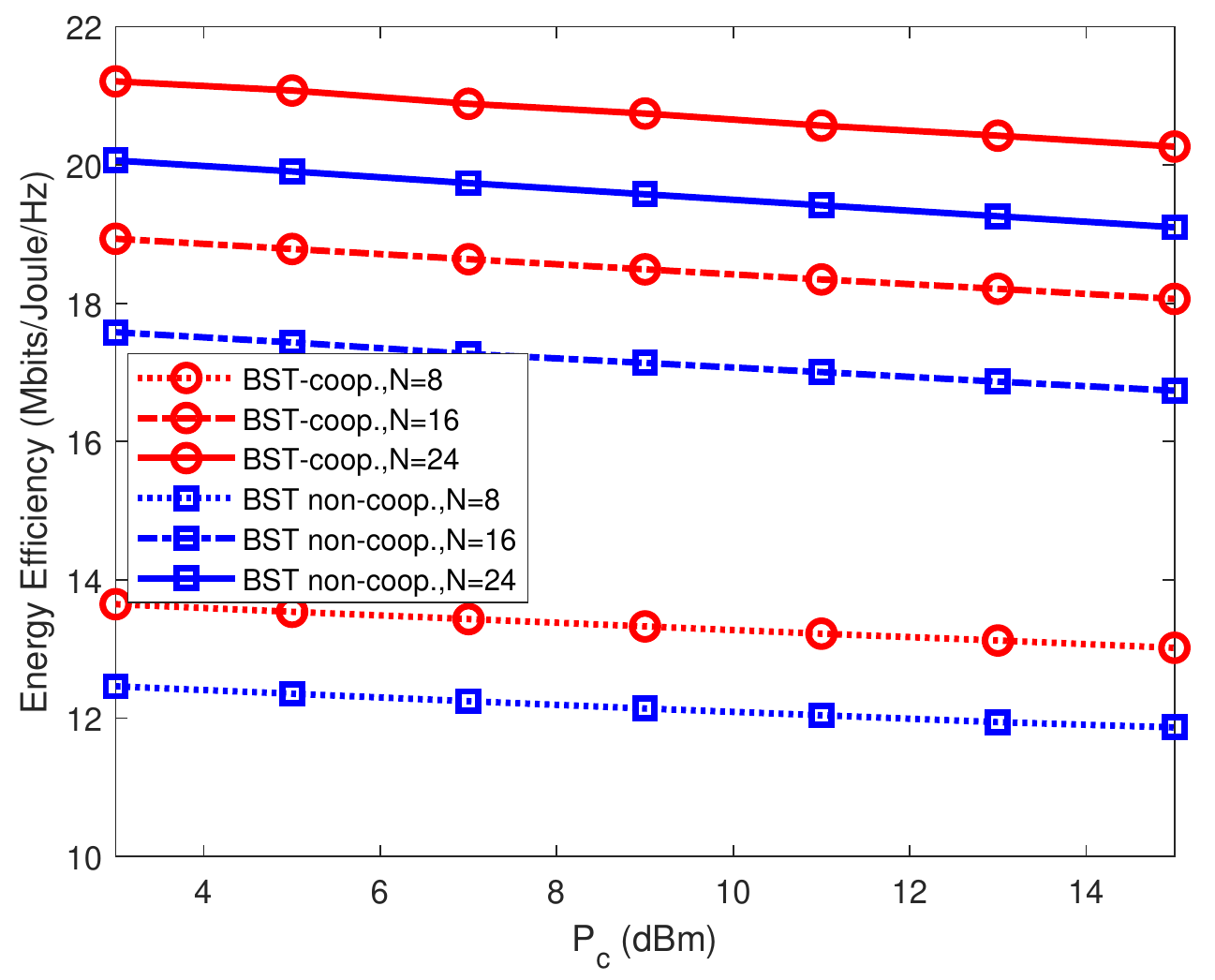}
 	\caption{The effect of increasing $P_{c}$ on energy efficiency under different number of antennas at the backscatter tag }
 	\label{f5}
 \end{figure}

Further, the impact of increasing values of circuit power $P_{c}$ on the energy-efficiency of the system under a different number of antennas equipped at the BST has been depicted in  Fig. \ref{f5}. It can be seen that the energy-efficiency decreases by increasing the value of circuit power. This is because an increment in the value of $P_c$ increases the total power consumption which ultimately decreases the energy-efficiency of the considered system. Further, simulation results evince the proposed BST-coop NOMA system outperforms its BST non-coop competitor in terms of energy-efficiency for different number antennas at the backscatter tag.  
   
\begin{figure}[t]
	\centering
	\includegraphics [width=0.48\textwidth]{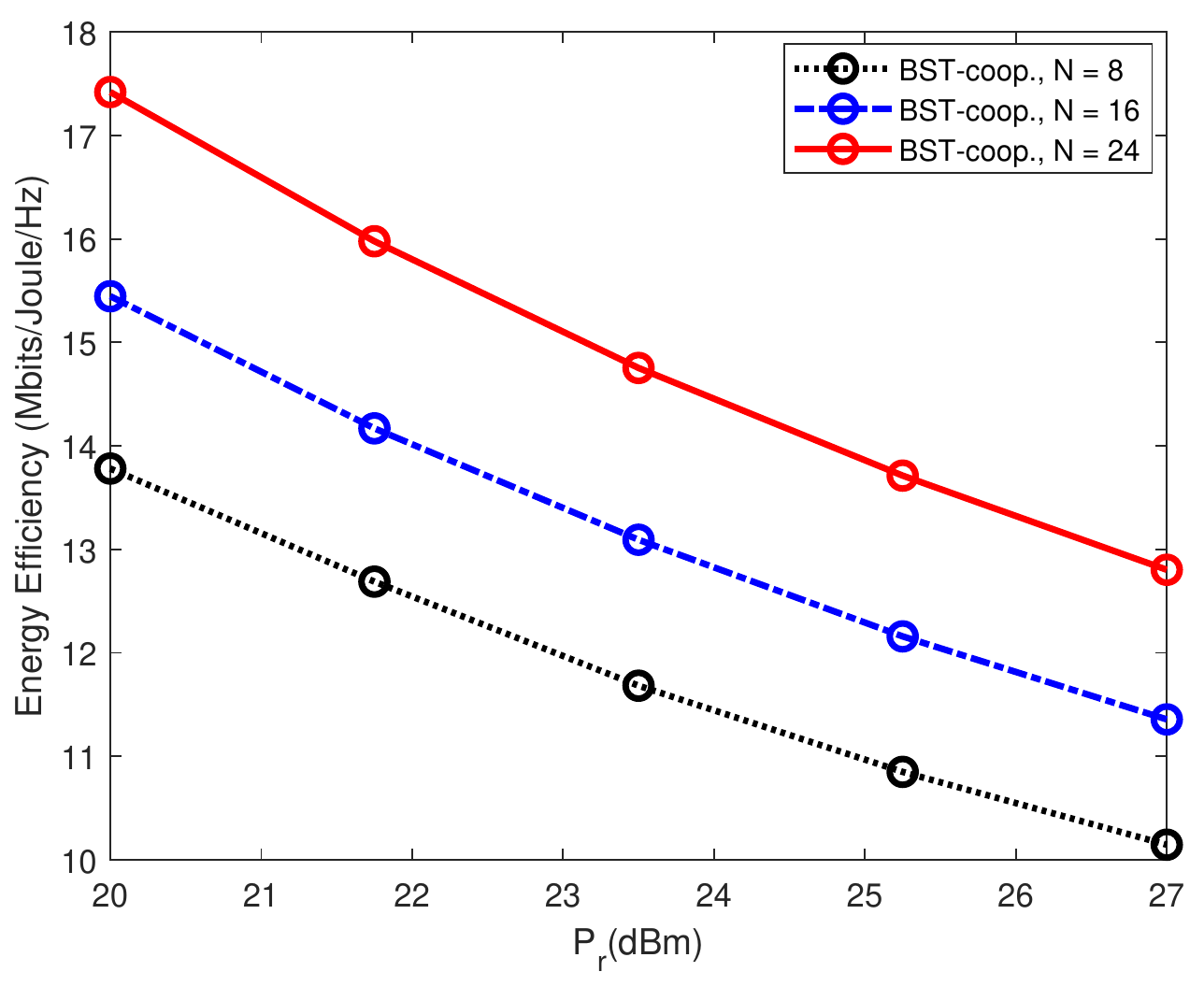}
	\caption{The impact of increasing values of relay-power $P_{r}$ on energy efficiency under different number of BST antennas}
	\label{f6}
\end{figure}

Finally, the effect of increasing power of relay nodes in each cluster of considered BST-assisted multi-clustered cooperative NOMA system has been investigated in Fig. \ref{f6}. Simulation results reveal that the energy-efficiency decreases by increasing the value of relay-power in each cluster. This is because by increasing the value of $P_{r}$ after a certain point, the total power consumption becomes more sensitive as compared to sum-rate which ultimately decreases the energy-efficiency of the considered system.  
   
\section{Conclusion and Remarks}
An energy-efficient alternating optimization framework has been proposed to maximize the EE for the considered BST-assisted cooperative IoT NOMA network, where the energy-efficiency of the system has been maximized by optimizing the active-beamforming vector $\mathbf{w}_k$ and PAC $\alpha_{k,n}$, $\alpha_{k,f}$ at the transmitter, as well as the passive-beamforming $\mathbf{g}$ at the backscatter tag. The proposed alternating optimization algorithm tackles the considered optimization problem in two stages. In the $1^{st}$ stage, ZF-based active-beamforming along with an efficient clustering technique has been exploited to combat the effect of ICI. Further, the intra-cluster interference is tackled by exploiting an efficient power-allocation policy that computes the PAC of IoT NOMA users under the desired constraints. In the $2^{nd}$ stage, by exploiting DC programming and SCA approximation, the considered non-convex passive-beamforming optimization problem has been transformed into a standard SDP problem, where Rank-1 solution of passive-beamforming vector is obtained based on the penalty-based method. Finally, the simulation results evince that the proposed alternating optimization algorithm exhibits an efficient energy-efficiency performance by achieving convergence within only a few iterations.

\bibliographystyle{IEEEtran}
\bibliography{Wali_Ref}

\begin{IEEEbiography}
	[{\includegraphics[width=1in,height=1.5in,clip,keepaspectratio]{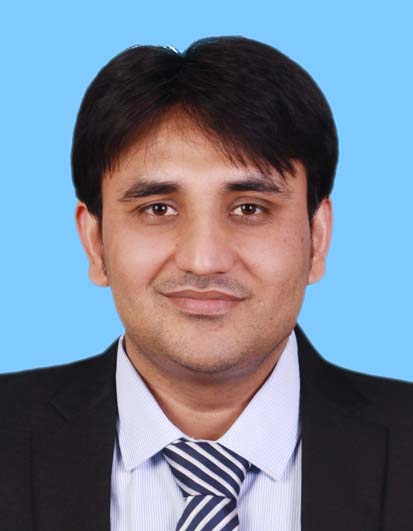}}]{MUHAMMAD ASIF} 
	was born in Rahim Yar Khan, Bahawalpur Division, Pakistan, in 1990. He received the Bachelor of Science (B.Sc) degree in Telecommunication Engineering from The Islamia University of Bahawalpur (IUB), Punjab, Pakistan, in 2013, and Master degree in Communication and Information Systems from Northwestern Polytechnical University (NWPU), Xian, Shaanxi, China, in 2015. He also received Ph.D. degree in Information and Communication Engineering from University of Science and Technology of China (USTC), Hefei, Anhui, China in 2019. Currently, Dr. Asif is working as a post-doctoral researcher at the Department of Electronics and Information Engineering in Shenzhen University, Shenzhen, Guangdong, China. He has authored/co-authored several journal and conference papers. His research interests include Wireless Communication, Channel Coding, Coded-Cooperative Communication, Optimization and Resource Allocation, Backscatter-Enabled Wireless Communication, IRS-Assisted Next-generation IOT Networks.
\end{IEEEbiography}

\begin{IEEEbiography}
	[{\includegraphics[width=1in,height=1.25in,clip,keepaspectratio]{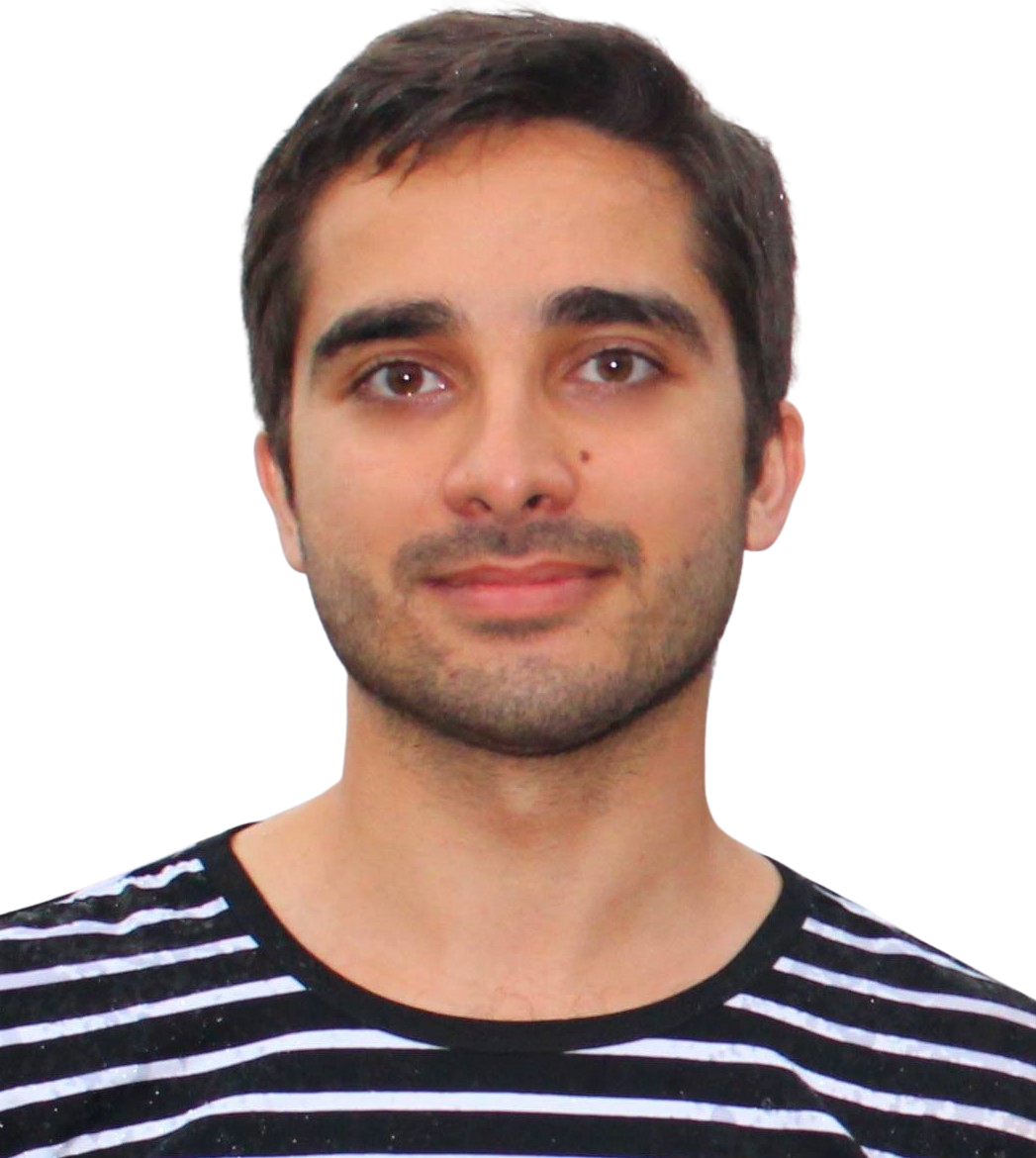}}]{Asim Ihsan}
	received the master’s degree in information and communication engineering from Xian Jiaotong University, Xi’an, China, and the Ph.D.degree in information and communication engineer-ing from Shanghai JiaoTong University, Shanghai,China. He is currently working as a Post-Doctoral Research Officer with the School of Computer Science and Electronic Engineering, Bangor University,U.K. He is also a Global Talent Visa Holder of U.K. His research interests include energy-efficient resource allocations for beyond 5G wireless communication technologies through convex/non-convex optimizations and machine learning.
\end{IEEEbiography}

\begin{IEEEbiography}
	[{\includegraphics[width=1in,height=1.5in,clip,keepaspectratio]{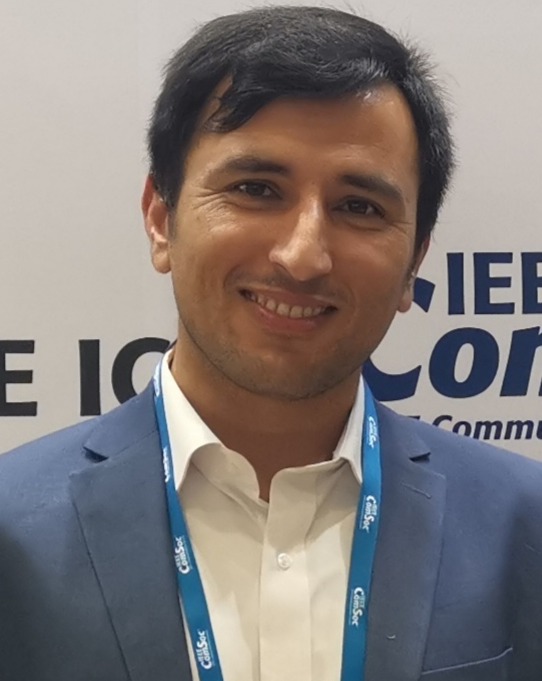}}]{Wali Ullah Khan} (Member, IEEE)
	received the Master degree in Electrical Engineering from COMSATS University Islamabad, Pakistan, in 2017, and the Ph.D. degree in Information and Communication Engineering from Shandong University, Qingdao, China, in 2020. He is currently working with the Interdisciplinary Centre for Security, Reliability and Trust (SnT), University of Luxembourg, Luxembourg. He has authored/coauthored more than 50 publications, including international journals, peer-reviewed conferences, and book chapters. His research interests include convex/nonconvex optimizations, non-orthogonal multiple access, reflecting intelligent surfaces, ambient backscatter communications, Internet of things, intelligent transportation systems, satellite communications, physical layer security, and applications of machine learning.
\end{IEEEbiography}

\begin{IEEEbiography}
	[{\includegraphics[width=1in,height=1.25in,clip,keepaspectratio]{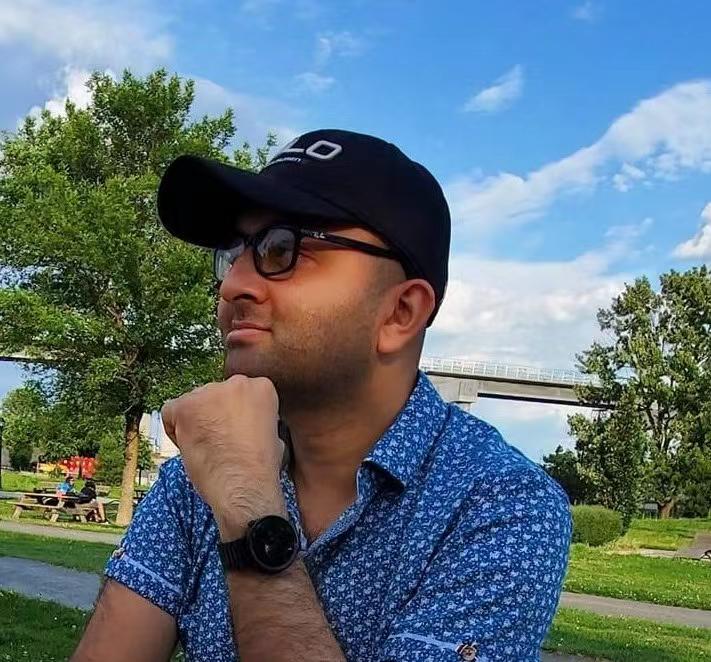}}]{Ali Ranjha}
	received his Ph.D. degree at Ecole de Technologie Superieure (ETS), Universite du Quebec, Montreal, Canada in January 2022, where he is also currently pursuing his postdoctoral research. In 2018, he completed his M.S. degree in innovation in telecommunications from Lancaster University, U.K. under a prestigious Higher Education Funding Council of England (HEFCE) bursary, His research interests span diverse areas, such as fundamental communication theory, unmanned aerial vehicle (UAV) communications, internet of things (IoT), ultra-reliable and low latency communications (URLLC), and optimization in resource-constrained networks.
\end{IEEEbiography}

\begin{IEEEbiography}
	[{\includegraphics[width=1in,height=1.25in,clip,keepaspectratio]{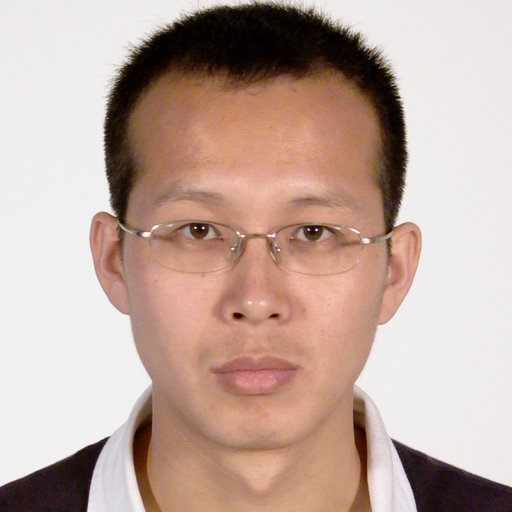}}]
	{Shengli Zhang} (Senior Member, IEEE) received the B.Eng. degree in electronic engineering and the
	M.Eng. degree in communication and information engineering from the University of Science and Technology of China, Hefei, China, in 2002 and 2005, respectively, and the Ph.D. degree from the department of Information Engineering, Chinese University of Hong Kong, Hong Kong, in 2008. He joined the Communication Engineering Department, Shenzhen University, Shenzhen, China, where he is currently a Full Professor. From March 2014 to March 2015, he was a Visiting Associate Professor with Stanford University, Stanford, CA, USA. He is the pioneer of Physical-Layer Network Coding. He has published several IEEE top journal papers and ACM top conference papers, including IEEE journal on selected areas in communications, IEEE transactions on wireless communications, IEEE transactions on mobile computing,
	IEEE transactions on communications, and ACM mobicom. His research interests include blockchain, physical layer network coding, and
	wireless networks. Prof. Zhang severed as an Editor for IEEE transactions on vehicular technology, IEEE wireless communications letters, and IET communications. He has also severed as a TPC member in several IEEE conferences.
\end{IEEEbiography}

\begin{IEEEbiography}
	[{\includegraphics[width=1in,height=1.25in,clip,keepaspectratio]{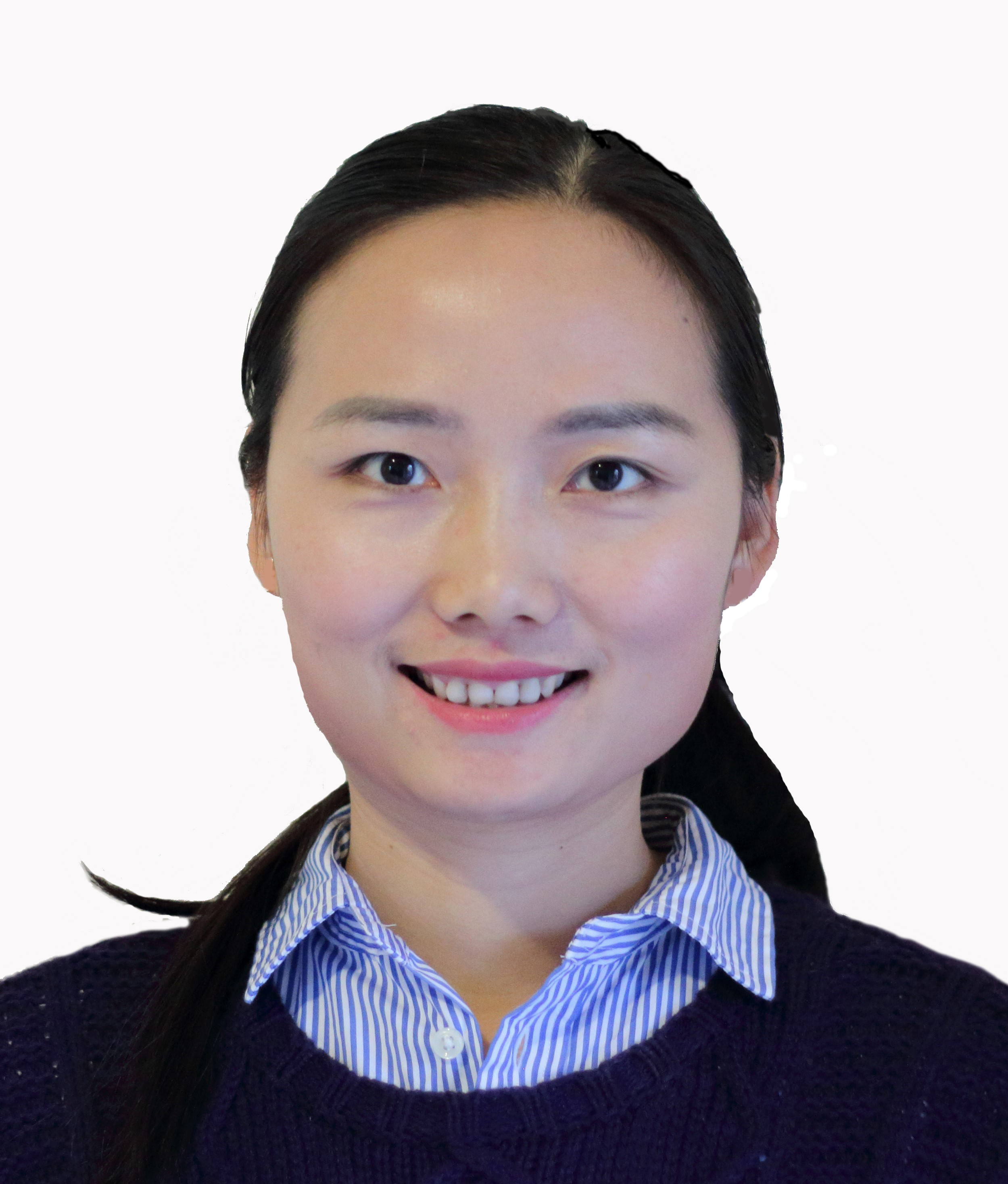}}]
	{Sissi Xiaoxiao Wu} (Member, IEEE) received the B.Eng. degree in electronic information engineering from the Huazhong University of Science and Technology, Wuhan, China, in 2005, the M.Phil. degree from the Department of Electronic and Computer Engineering, Hong Kong University of Science and Technology, Hong Kong, in 2009, and the Ph.D. degree in electronic engineering from the Chinese University of Hong Kong (CUHK), Hong Kong, in 2013.,From December 2013 to November 2015, she was a Postdoctoral Fellow in the Department of Systems Engineering and Engineering Management, CUHK. From December 2015 to March 2017, she was a Postdoctoral Fellow in the Signal, Information, Networks and Energy Laboratory supervised by Prof. A. Scaglione of Arizona State University, Tempe, AZ, USA. She is now an Associate Professor at the Department of Communication and Information Engineering, Shenzhen University, Shenzhen, China. Her research interests are in wireless communication theory, optimization theory, stochastic process, and channel coding theory, and with a recent emphasis on the modeling and data mining of opinion diffusion in social networks. She is now an Associate Editor of IEEE Transactions on Vehicular Technology and serving as an IEEE Signal Processing Society SAM Technical Committee Member.
\end{IEEEbiography}

\end{document}